Published in *Light: Science & Applications* (2026)

DOI

https://doi.org/10.1038/s41377-025-02174-5

Januariyasa, I.K., Reda, F., Liubimtsev, N. *et al.* Stress-driven photo-reconfiguration of surface microstructures via vectorial field-guided lithography. *Light Sci Appl* **15**, 194 (2026). https://doi.org/10.1038/s41377-025-02174-5.

# Stress-driven photo-reconfiguration of surface microstructures via vectorial field-guided lithography

I Komang Januariyasa,[1] Francesco Reda,[1] Nikolai Liubimtsev,[2] Pawan Patel,[2,3] Cody Pedersen,[4] Fabio Borbone,[5] Marcella Salvatore,[1] Marina Saphiannikova,[2,3,*] David J. McGee,[4,*] and Stefano Luigi Oscurato[1,*]

[1]Physics Department "E. Pancini", University of Naples Federico II, Complesso Universitario di Monte Sant'Angelo, via Cinthia, 80126, Naples, Italy.

[2]Division Theory of Polymers, Leibniz Institute of Polymer Research Dresden, 01069 Dresden, Germany.

[3]Faculty of Mechanical Science and Engineering, Dresden University of Technology, 01062 Dresden, Germany

[4]Department of Physics, The College of New Jersey, Ewing, New Jersey 08628, USA.

[5]Department of Chemical Sciences, University of Naples "Federico II", Complesso Universitario di Monte Sant'Angelo, Via Cintia, 80126 Naples, Italy.

[*]Stefano Luigi Oscurato, e-mail: stefanoluigi.oscurato@unina.it

[*]Marina Saphiannikova, e-mail: grenzer@ipfdd.de

[*]David J. McGee, e-mail: mcgeed@tcnj.edu

## Abstract

Pattern formation driven by mechanical stress plays a fundamental role in shaping structural organization in both natural and human-made systems. Using light as a vectorial stimulus may offer a powerful route to control stress-induced pattern formation in materials. However, achieving localized, programmable, and predictable control of individual microstructures via structured polarization fields has remained a major challenge. Here, we introduce vectorial field-guided lithography, a novel approach that leverages fully structured polarization fields as lithographic tools to enable the stress-driven reconfiguration of pre-patterned azopolymer microstructures with an unprecedented degree of flexibility, complexity, and diversity. By building on the Viscoplastic PhotoAlignment model, which describes the azopolymer deformation as stress response to structured light, we quantitatively demonstrate and predict complex surface architectures generated by programmable light-induced stress pathways using a digital polarization rotator implemented via a spatial light modulator. We model and experimentally achieve single-step formation of anisotropic, bent, and chiral microstructures from a single pre-patterned geometry. Our results reveal an exceptional control over local microstructure morphology and establish, for the first time, a comprehensive theoretical framework capable of quantitatively designing and fabricating target morphologies on azopolymers. This work moves beyond conventional intensity-based photopatterning and demonstrates that the full vectorial nature of light can dictate the mechanical reshaping of functional polymer surfaces, providing a new platform for the programmable design of complex micro-architectures with applications in photonics, microfluidics, and biology.

# Introduction

Stress-driven pattern formation is a fundamental mechanism that governs the emergence of complex structures across spatial and temporal scales in both natural and engineered systems. On Earth's surface, tectonic stress sculpts mountains and fault lines.[1] In living organisms, differential growth induces mechanical instabilities, leading to the characteristic folds of the human brain,[2] the wrinkling of the skin,[3] and the morphological features in plant leaves.[4] Even at the microscopic level, bacterial colonies self-organize into complex morphologies under the influence of mechanical stress.[5] Despite their diverse origins, these systems share a common underlying principle. When internal stress accumulates, the system undergoes morphological reorganization to minimize free energy, often through symmetry breaking, giving rise to structured topographies. Inspired by this universal process, the engineering of controlled stress-driven deformations has emerged as a powerful strategy to design functional surfaces. Examples include tunable wrinkling in soft materials,[6] origami-inspired folding,[7] and buckling of stretchable materials.[8]

All of these examples leverage the relationship between a triggering external action and the stress generated within the material, with this interaction guiding the material into forming structures. Typically, these systems harness the stress induced by external stimuli such as temperature, humidity, or mechanical load.[9,10] Light, which allows remote, non-contact control with high spatial and temporal accuracy, offers distinct advantages among the other stimuli.[11–13] More importantly, light inherently carries vectorial information through its polarization state. This degree of freedom remains largely untapped for conventional stress-driven surface structuring and light-based photopatterning processes, such as standard optical lithography.

Amorphous azobenzene-containing polymers (referred to as azopolymers), with their unique light-induced anisotropic stress-driven deformation, offer a route to bridge this gap and harness the vectorial nature of light for surface patterning.[14–16] Key elements for this process are the photochemical properties of the azobenzene chromophores embedded in the hosting polymeric material matrix. Following cyclic *trans-cis* photoisomerization with UV/visible light, the azobenzene molecules reorient perpendicular to the direction of the local light polarization,[14,15] transferring their alignment to the polymer to which they are attached through either covalent[17,18] or supramolecular interactions.[19,20] The mechanical stress resulting from the photoalignment is relieved by directional plastic deformation of the material in the irradiated area.[18,21] When the azopolymer is prepared as a thin film, this light-induced phenomenon directly produces relief patterns on the free surface of the film. The morphology of such reliefs reflects the local geometry of the incident vectorial light field, showing a dependence on both the spatial intensity distribution[22–24] and its local polarization state.[25–27] Empirical and phenomenological exploitation of the vectorial nature of the light-induced surface deformation of azopolymers has been used to fabricate a wide range of different surface relief architectures. Interestingly, complex anisotropic microstructures, with tailored morphology even in three dimensions, can also be obtained when the intensity or the polarization of the incident light is used to reconfigure the geometry of already pre-patterned surfaces. Complex isometric,[28] and strongly unidirectional[29–31] and bidirectional[32–34] anisometric architectures have been achieved from pristine arrays of azopolymer micropillars in this scheme.

The modeling of azopolymer photopatterning processes has been a long-standing challenge due to the complex interplay between light parameters and material response.[15] However, a recent theoretical framework, the Viscoplastic PhotoAlignment (VPA) model, has been proposed to describe the process as a unique stress-driven vectorial phenomenon initiated by the light. This picture could ultimately bridge the existing gap between the nanoscopic photomechanics of azobenzene molecules and the resulting macroscopic surface relief geometries.[18] Beyond rationalizing the complex multiscale photoinduced dynamics in these materials, the relevance of this model lies in the possibility of finally having predictive and inverse design capabilities for the surface reliefs on azomaterials. In addition, extending and validating the model to the general case of irradiation with structured vectorial light fields could provide foresight for the design of unconventional illumination schemes and materials for new lithographic processes.

In this work, we demonstrate a new level of stress-driven morphological control based on the programmable photo-reconfiguration of pre-patterned azopolymer surfaces using structured vectorial light fields. Specifically, we analyze the photo-reconfiguration of azopolymer micropillars illuminated by spatially varying polarization patterns, which are generated by a digital polarization rotator implemented via a computer-controlled spatial light modulator.[35,36] This optical system offers large flexibility in the polarization pattern design, making our configuration a prototype of a more general vectorial field generator able to control the stress-driven morphological deformation of the azopolymers, with an exceptional degree of complexity and diversity. Furthermore, our experimental framework provides an ideal test ground to further validate the VPA model for arbitrary structured light. The combined experimental and theoretical study presented here demonstrates unprecedented quantitative, predictive capabilities on the morphology and the stress-driven deformation dynamics of complex azopolymer microstructures. Our results establish the concept of the vectorial field-guided lithography, summarized in Fig. 1, whereby a rationally designed vectorial light pattern is directly converted into programmable microstructure morphologies via tailored light-induced stress pathways. The presented framework opens up the exploration of vectorial external stimuli as a tool for shaping even diverse material systems, paving the way for applications in different fields such as photonics, microfluidics, and biology.

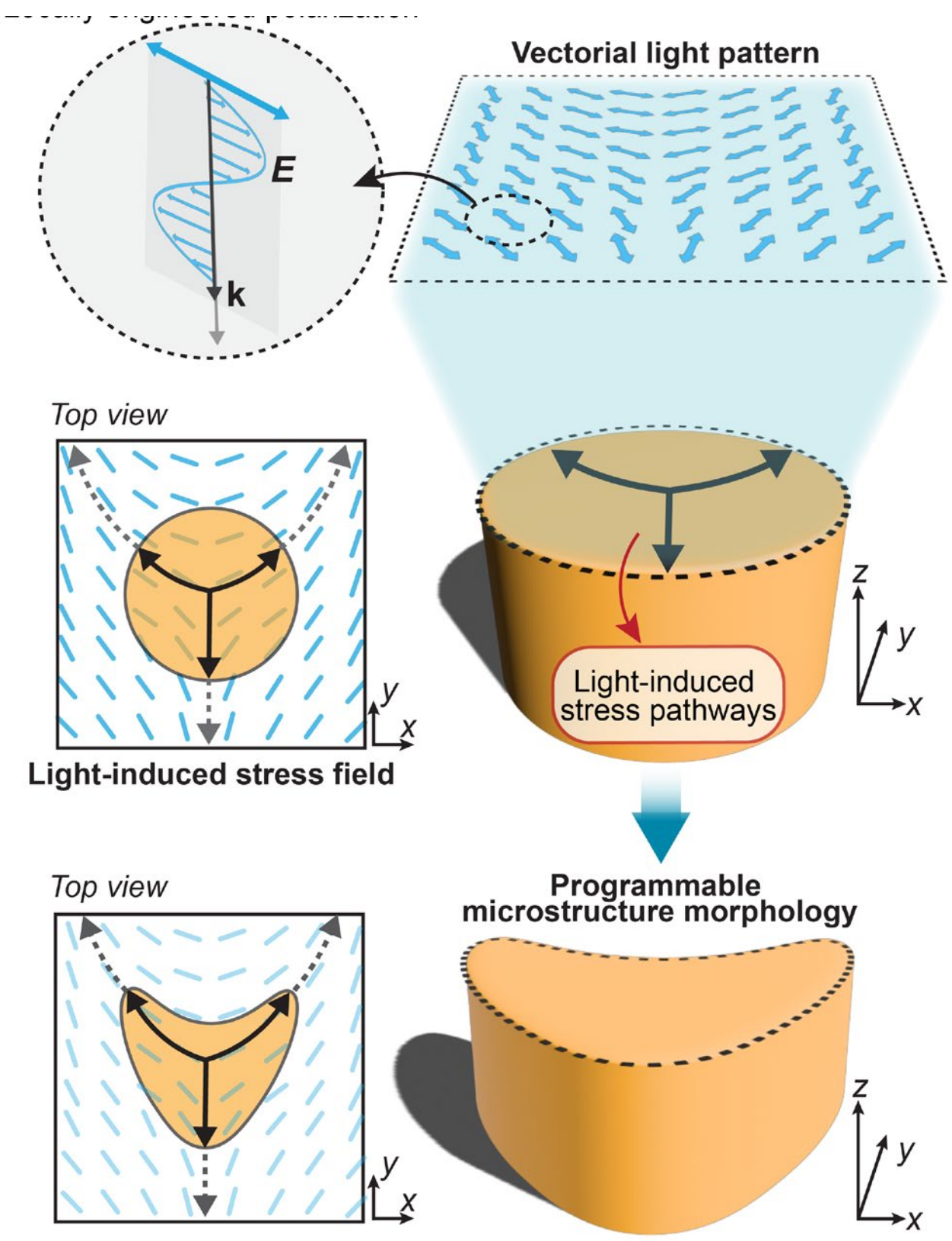

**Fig 1. Concept of vectorial field-guided lithography in azomaterial microstructures.** A vectorial light field with locally engineered linear polarization directions (first row) is irradiated onto the top surface of a cylindrical micropillar, inducing stress pathways (black arrows in the middle row). Their directions depend on the light field distribution (dashed blue lines) and guide the anisotropic structural deformation of the illuminated micropillar along the same directions (bottom row).

# Results

## 1. Stress-driven morphological deformations from linearly polarized light

Before introducing the general framework of the vectorial field-guided lithography, we first rationalize the photomechanical response of azopolymer micropillars to uniform linearly polarized light as a baseline of the stress-driven morphological deformation described by the VPA model. This photo-reconfiguration scenario,[30,34] schematized in Fig. 2a, has been used to induce either directional small, reversible deformations[37] or complete remorphing of azopolymer micropillar arrays.[38] However, previous studies have been largely limited to empirical observations and phenomenological accounts of polarization-driven deformation, lacking a quantitative, predictive framework.

For our analysis, the azomaterial surface is designed as an array of cylindrical micropillars and placed in the x-y plane. The surface photo-deformation involves illumination with a homogeneous laser beam, which in our scheme is linearly polarized along the x-axis. The polarization direction is indicated by the blue arrow identifying the direction of electric field (**E**) in Fig. 2a. A uniform intensity distribution is taken into account here to minimize contributions from intensity gradients, which are known to be an additional relevant mechanism driving surface deformation in azopolymer films[22,23] and microstructures.[39] Moreover, the uniform illumination also ensures that each micropillar within the irradiated region experiences the same photo-deformation dynamics.

Fig. 2b shows the Scanning Electron Microscopy (SEM) images of our fabricated pristine structured surface. For the photo-reconfiguration experiment, a linearly polarized laser beam at a wavelength of 488 nm was used. The beam was sufficiently expanded to ensure spatial uniformity of the intensity in the illuminated area. Details on our azopolymer, the preparation and the geometry of the pre-patterned array, and the optical setup are presented in Materials and methods and in Fig. S1-S2-S3.

Fig. 2c shows the reconfiguration process of a micropillar in the array achieved by exposing the surface for 18 minutes to a laser intensity of 0.16 W/cm². According to the expected phenomenology,[30] the irradiation breaks the symmetry of the initially isotropic cylindrical micropillar, resulting in an elongated microstructure in the direction of the polarization of the incident light beam. The same happens for all the micropillars in the homogeneously illuminated area.

In the VPA model, the amorphous polymer is schematized as a series of rod-like polymer backbone segments, to which the azobenzene chromophores are attached as side chains, preferentially at right angles. The directional photo-deformation of the azopolymer is triggered by the light-induced reorientation of the azobenzene chromophores into the plane perpendicular to the electric field **E**. Since the azopolymer is in a glassy state at room temperature, the VPA model assumes a rigid coupling between the azobenzene chromophores and the polymer backbones. This coupling forces the backbones to align along the polarization direction (Fig. 2d).[18,21] In this process, the electric field vector **E** of the homogeneously polarized beam plays the role of a nematic director that breaks the symmetry of the initially isotropic molecular system.[15] Importantly, the emergence of orientation anisotropy within the polymer matrix induces mechanical stress; consequently, the alignment of the polymer backbones generates a light-induced stress.

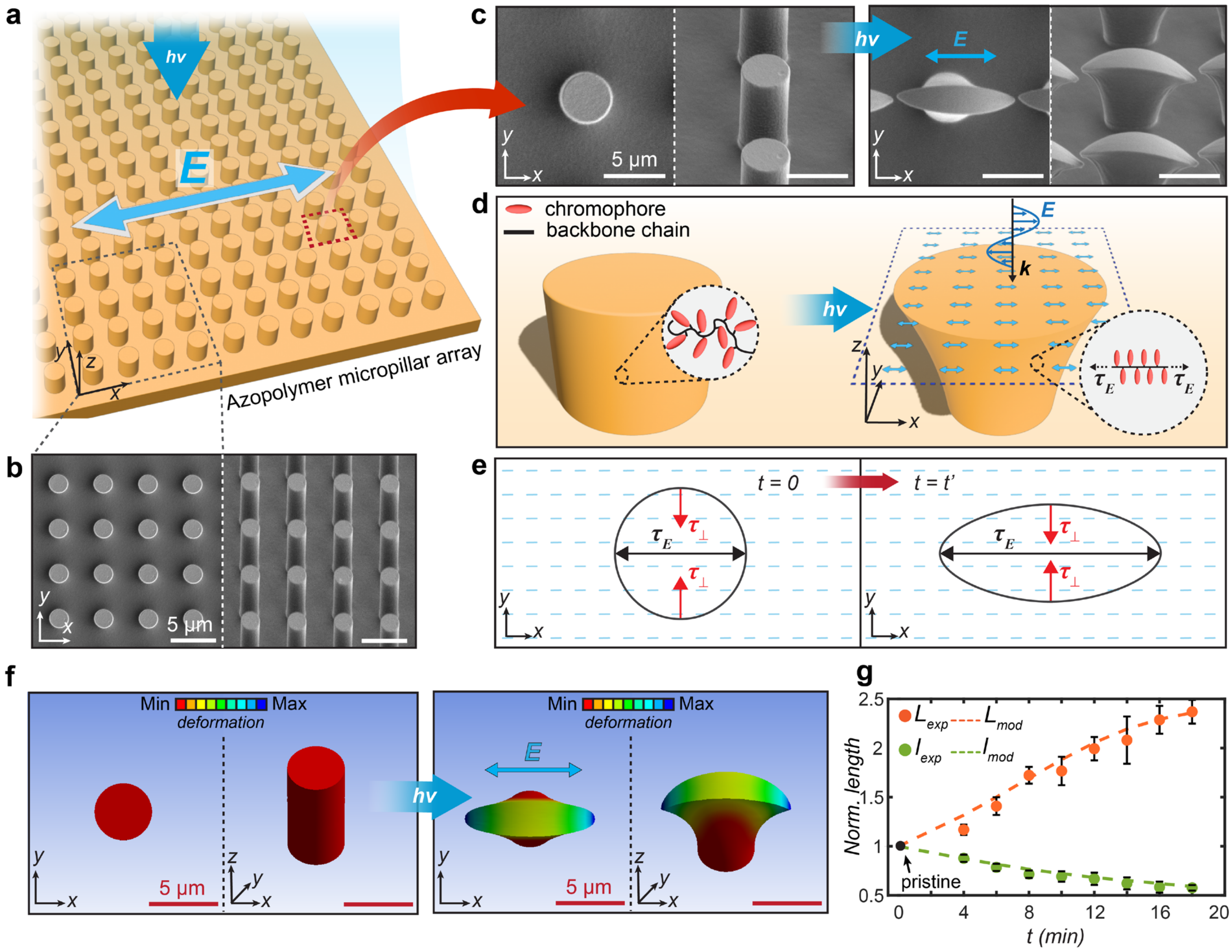

**Fig 2. Encoding of polarization direction to morphological anisotropy. a)** Illumination scheme of homogeneous linearly polarized beam onto an azopolymer micropillar array. **b)** SEM images of the pristine state of the micropillar array from top (left) and tilted (right) views. **c)** SEM images of a single pristine micropillar from top and side views (left) and of an anisotropically deformed micropillar resulting from illumination with uniformly polarized light. **d)** Illustration of the relationship between molecular photoalignment and the micro-scale photo-deformation in the VPA model. **e)** Scheme of the emerging stress induced by the photoalignment at the very initial stages of illumination ($t = 0$) and after a certain exposure time ($t = t'$). The tensile stress ($\tau_E$) and the compressive stress ($\tau_\perp$) are parallel and orthogonal to the polarization direction, respectively. **f)** VPA modeled morphological evolution of the azopolymer micropillar (top and tilted view) corresponding to the experimental results in panel c). The colormap represents vertical displacement: red corresponds to the initial state, while dark blue indicates the maximum downward displacement. **g)** Dynamic evolution of the experimental and modeled deformation of the micropillars. $L_{exp}$, $l_{exp}$, are the lengths of major and minor axes of the top surface measured in the experiment, while $L_{mod}$, and $l_{mod}$ are the same lengths extracted from the model. The geometrical parameters have been normalized by the diameter ($D_0$) of the pristine micropillar. Error bars indicate the standard deviation of 10 measured pillars in the array for each exposure time.

The complete formalism linking the kinetic equations of azobenzene isomerization to the stress induced in the polymer is detailed in the Supplementary note 1. To emphasize the most relevant aspect of the VPA model, we derived an insightful expression that directly relates the light-induced stress tensor $\boldsymbol{\tau}$ to the polarization direction $\hat{\mathbf{E}}=\mathbf{E}/|\mathbf{E}|$ of linearly polarized light:

$$\boldsymbol{\tau} = \tau_0\left(\hat{\boldsymbol{E}}\hat{\boldsymbol{E}} - \boldsymbol{\delta}/3\right). \tag{1}$$

In this relation, $\tau_0$ is the magnitude of the stress, which increases with the light intensity,[40] the dyadic product $\hat{\boldsymbol{E}}\hat{\boldsymbol{E}}$ is a tensor that contains the information about the principal axes of stress tensor, and $\boldsymbol{\delta}$ is the unit tensor. This relation was derived assuming an initial isotropic state of the polymer backbones and is valid for the initial stages of irradiation. The main feature emerging from Equation (1) is that the major principal axis of the stress tensor for linearly polarized light is aligned with the local director $\hat{\boldsymbol{E}}$, whose uniform spatial distribution is represented by the blue segments in Fig. 2e. In particular, for the linear polarization $\hat{\boldsymbol{E}} = (1,0,0)$ aligned with the x-axis, $\boldsymbol{\tau}$ explicitly becomes:

$$\boldsymbol{\tau} = \tau_0\begin{pmatrix} 2/3 & 0 & 0 \\ 0 & -1/3 & 0 \\ 0 & 0 & -1/3 \end{pmatrix}. \tag{2}$$

According to Equation (2) the stress tensor is represented by three principal normal stresses: the tensile component $\tau_E = 2\tau_0/3$ acting along the major principal direction (x-axis here), and the compressive components $\tau_\perp = -\tau_0/3$ acting in the two perpendicular minor principal directions (Fig. 2e).[41] Such a light-induced stress tensor leads to the uniaxial elongation of the illuminated azopolymer along the local light polarization direction and to the compression along the orthogonal directions. Following this description, the stress field of a uniform polarization pattern can be illustrated as a single stress pathway (the black arrow in Fig. 2e), corresponding to the major principal direction of the system, along which the polymer deformation mainly occurs.

It is worth noting that the results of the VPA model apply to any polarization state of the irradiated field. In particular, for an elliptically polarized light, relation (1) predicts a general biaxial deformation of the azopolymer micropillar, where two principal stress directions correspond to the axes of the light polarization ellipse.[42,43]

To model the micropillar photo-deformation in the situation of Fig. 2, we performed finite element simulations using the complete VPA model (Supplementary note 1), where the azopolymer is treated as a viscoplastic material, and the light-induced stress field was implemented via a custom subroutine in ANSYS software (see Materials and methods). As discussed in detail in a recent review,[15] the main physical mechanism underlying the model is the plasticization of glassy azopolymers under light-induced stress (1), whose magnitude $\tau_0$ has been shown to exceed the yield stress $\tau_{yield}$ by a large margin. This observation justifies the use of a viscoplastic material model in the present work. Other possible mechanisms, such as pronounced photosoftening or a photoinduced solid-to-liquid transition arising from a reduction of the glass transition temperature,[44] can be safely excluded for the class of azopolymers employed in this study.

Fig. 2f shows the modeled initial azopolymer microstructure and its morphology at the end of the photo-deformation process. The anisotropic microstructure that results from the VPA model is clearly in excellent agreement with experimental micropillars in Fig. 2c.

To further investigate the agreement beyond just the morphology at the final stages of the reconfiguration process, we analyzed the temporal dynamics of the azopolymer photo-deformation, extracted from the characterization of the microstructure morphology at different exposure times. The micropillar deformation, both in experiments and model, is quantitatively described by measuring the major ($L$) and the minor ($l$) axis of the pseudo-elliptical shape of the top surface, respectively. Both parameters were normalized to the diameter ($D_0$) of the pristine circular micropillar profile before exposure (see also Movie S1 and Fig. S4 for dynamic modeled and experimental evolution, respectively). The comparison, reported in Fig. 2g, demonstrates quantitative agreement at each stage of the deformation process.

## 2. Programmable optical field-guided deformations

The analysis conducted above for the uniform linear polarization state allows us to extend the established concept of the uniform stress pathway of the VPA model to the description of deformation induced by vectorial polarization states. We limit our treatment to spatially structured patterns of linear polarization, but the results are generalizable to arbitrary polarization states. Namely, according to Equation (1), when the polarization field is spatially structured, (i.e. $\mathbf{E} = \mathbf{E}(\mathbf{r})$) the stress field also becomes spatially varying ($\boldsymbol{\tau} = \boldsymbol{\tau}(\boldsymbol{r})$). In this condition, at each point $\mathbf{r}$ within the irradiated azopolymer, the local major principal axis of the stress aligns with the direction of the local linear polarization, enabling a spatially controlled deformation ultimately directed by the optical field.

The vectorial dependence of the stress-driven deformation can be easily visualized through the array of azopolymer micropillars, illuminated with discretized structured patterns of linearly polarized light. In this configuration, each microstructure in the array is deformed according to the local linear polarization direction, acting as a morphological polarization sensor. For our experiments, we used a digital polarization rotator based on a Spatial Light Modulator (SLM) to irradiate the pre-patterned azopolymer micropillar array. This optical system is designed to rotate the azimuth angle $\varphi$ of a linearly polarized beam and consists of an SLM interposed between two quarter-wave plates whose axes are each oriented at ±45° with respect to the incident linear polarization (see Materials and methods).[20,36] The output polarization angle $\varphi$ is determined by the phase delay introduced between the two orthogonal components of the circularly polarized light incident on the device, according to the gray level of a computer-generated image displayed on the SLM (see also Fig. S5).[25,45] The digital image can be designed to create spatially varying maps $\varphi = \varphi(x, y)$ of linear polarization by assigning an independent gray value to each pixel of the SLM, as schematically shown in Fig. 3a.

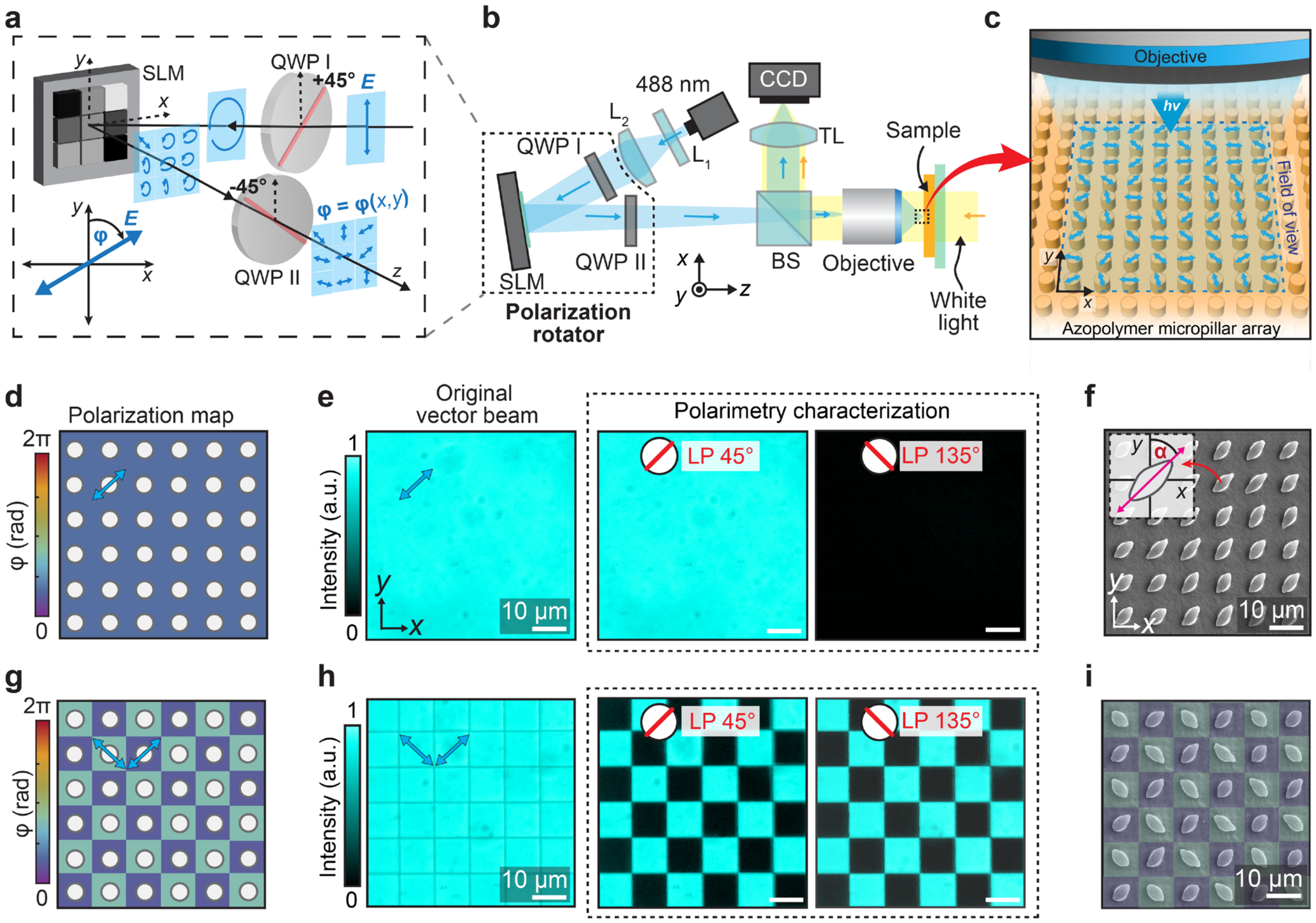

**Fig 3. Generation of engineered vector beams and spatial anisotropic control of the photo-deformation.** **a)** Working principle of a digital polarization rotator based on the spatial light modulator (SLM). The fast axes of first (QWP I) and second (QWP II) waveplates are oriented at +45° and -45° relative to the *y*-axis, respectively. The polarization azimuth angle $\varphi$ is defined as the angle between the polarization direction and the *y*-axis along the clockwise rotation direction. **b)** Optical setup scheme. The digital polarization rotator is integrated into a projection system that directs the illumination onto the azopolymer micropillar surface using a 40× objective. White light propagating from the back of the sample, and a CCD camera are used to observe the surface of the sample. **c)** Illustration of the projected spatially-varying vector beam on the micropillar array. **d)** Uniform polarization map designed at $\varphi = +45°$. **e)** Polarimetry characterization for experimental confirmation of the targeted polarization state designed in d). The optical intensity image of the irradiated light pattern is collected at the sample plane after a linear polarizer with the polarization axis set to 45° (left), and 135° (right). **f)** SEM image of the deformed micropillar after the exposure with the light pattern shown in e).

**g)** Polarization map of checkerboard polarization pattern designed with $\varphi_1 = +45°$ and $\varphi_2 = -45°$. **h)** Experimental confirmation of the targeted polarization state designed in g) as described for panel e). **i)** SEM image of the deformed micropillars after the exposure with the light pattern shown in h).

For the 256 gray levels addressable to our SLM (8-bit images), the achievable angular resolution in the output polarization angle is ~1.4°, which allows fine control over the polarization orientation across the beam profile. In our experimental setup (Fig. 3b), the digital polarization rotator was integrated into an optical projection system (see Materials and methods for further details) to initiate the photo-deformation process of the azopolymer micropillar arrays with spatial selectivity (Fig. 3c). Such an optical configuration enables selective, site-specific remorphing of micropillars across the sample with polarization-dependent individual tuning.

Fig. 3d-f first demonstrates the ability of this system to replicate the experimental case presented in the previous section for homogeneous micropillar photo-deformation along on-demand directions. For this experiment, we targeted a uniform linear polarization illumination with an azimuth angle $\varphi = +45°$ across the entire field of view of the system (approximately 130 μm × 78 μm, see also Materials and methods). To enhance visualization here and in the remainder of this paper, we use a violet-to-red colormap to directly represent the designed polarization maps. The colors encode the polarization angle $\varphi$ emerging from the digital rotator system for each gray value addressed to the SLM according to the measured calibration curve reported in Fig. S5. Within this representation, the uniform polarization pattern for $\varphi = +45°$ appears as the homogeneous blue color in Fig. 3d.

Fig. 3e shows the experimental optical images representing the irradiated homogeneous light pattern, together with a polarimetry characterization confirming the achievement of the targeted polarization state as the output of the digital polarization rotator. It should be mentioned that, for the sake of improving the visualization, the optical images in Fig. 3 were collected through a CCD camera using an incoherent LED at 477 nm (with full-width-half-maximum of 29 nm) as a light source. In this way we could drastically reduce unwanted interference fringes caused by the coherent 488 nm laser source used in the photo-reconfiguration experiments (see Materials and methods), without affecting the main outcomes of the characterization of the digital polarization rotator system.

As expected from the previous analysis, a homogeneous elongation of all illuminated micropillars in the direction of the linear polarization was obtained in the experiment (Fig. 3f, and Table S1), with a measured average orientation angle $\alpha = +43° \pm 5°$, in full agreement with the target design.

We now demonstrate the next level of control enabled by our optical configuration, where spatially varying polarization maps can be designed to individually reshape the micropillars with independently engineered orientation $\alpha$ across the entire field of view. Owing to the square lattice arrangement of the micropillars in our array, the optical field of view can be easily segmented into square regions, each targeting an individual micropillar with a specific linear polarization orientation (see Materials and methods). The polarization map in Fig. 3g was designed as a checkerboard pattern of linearly polarized light oriented at alternating orthogonal azimuth angles of $\varphi_1 = +45°$ and $\varphi_2 = -45°$. Similarly to the previous uniform illumination, the optical images of the irradiated homogeneous light pattern, together with polarimetry characterization confirming the alternating targeted polarization state, are shown in Fig. 3h. In the image of the irradiated field (first image in Fig. 3h), distinct dark lines are visible at the edges of each of the square regions, as expected from the emergence of singularities at positions of sharp polarization variations in the optical wavefield.[46] These field discontinuities provide additional evidence of the system's ability to control vectorial light. However, they do not play a role in this photo-reconfiguration experiment as they are located outside to the targeted azopolymer micropillars.

The irradiation of the checkerboard polarization pattern on the azopolymer array resulted in the stress-driven regular reshaping of the micropillars, with alternating photo-deformation orientations measured from the SEM image (Fig. 3i) of $\alpha_1 = +43° \pm 4°$ and $\alpha_2 = -43° \pm 5°$, in full agreement with the designed polarization map. Additional experimental examples, highlighting the flexibility of the optical system and the peculiar alternating surface patterns achievable through the stress-driven photo-reconfiguration of this square micropillar array and the possibility to achieve progressive linear rotations, are shown in Fig S6 and S7, respectively.

To conclude with this section, it is worth highlighting that the framework proposed here can be further extended to enable coupled modulation of both light intensity and polarization by introducing an additional linear polarizer into the optical setup, as illustrated in Fig. S8-9. This extension provides the capability to simultaneously control both the pillar elongation direction (polarization) and the deformation strength (intensity) and could represent a first prototype of a full vectorial lithography able to achieve arbitrary complex directional patterns on surfaces.

## 3. Stress-driven complex microstructures from vectorial field

We describe now the complex reconfiguration of individual microstructures through locally engineered stress pathways originated by high-resolution structured polarization fields. This demonstrates the concept of maskless, programmable, and spatially-addressable vectorial lithography in its essential form.

Fig. 4a shows the principle underlying the design of this photo-reconfiguration scheme. Due to the high density of addressable pixels in the optical pattern projected onto the azopolymer surface ($\sim$15 pixels per $\mu m$), we can design high-resolution polarization maps that vary spatially across the surface of a single micropillar (diameter of $D_0 = 6.7\ \mu m$ for these experiments, see also Materials and methods). According to the VPA model, each volume element $dV$ of the irradiated azopolymer would experience the effect of the stress field generated by the local polarization state of the light (Fig 4a), tending to deform along the major principal direction of the maximum tensile component (with the same phenomenology described in Fig. 2). To have distinct directional deformations, adjoining volume elements should deform along closely varying directions that are only slightly different. This requires that the orientation $\varphi(x, y)$ of the light polarization field, which dictates the major principal stress direction (see also Supplementary note 2 and Fig. S10), varies smoothly across the x-y plane, as shown in Fig. 4a. In this condition, locally curved stress pathways emerge (black lines in Fig. 4b), which lead to complex deformation geometries for the azopolymer microstructures.

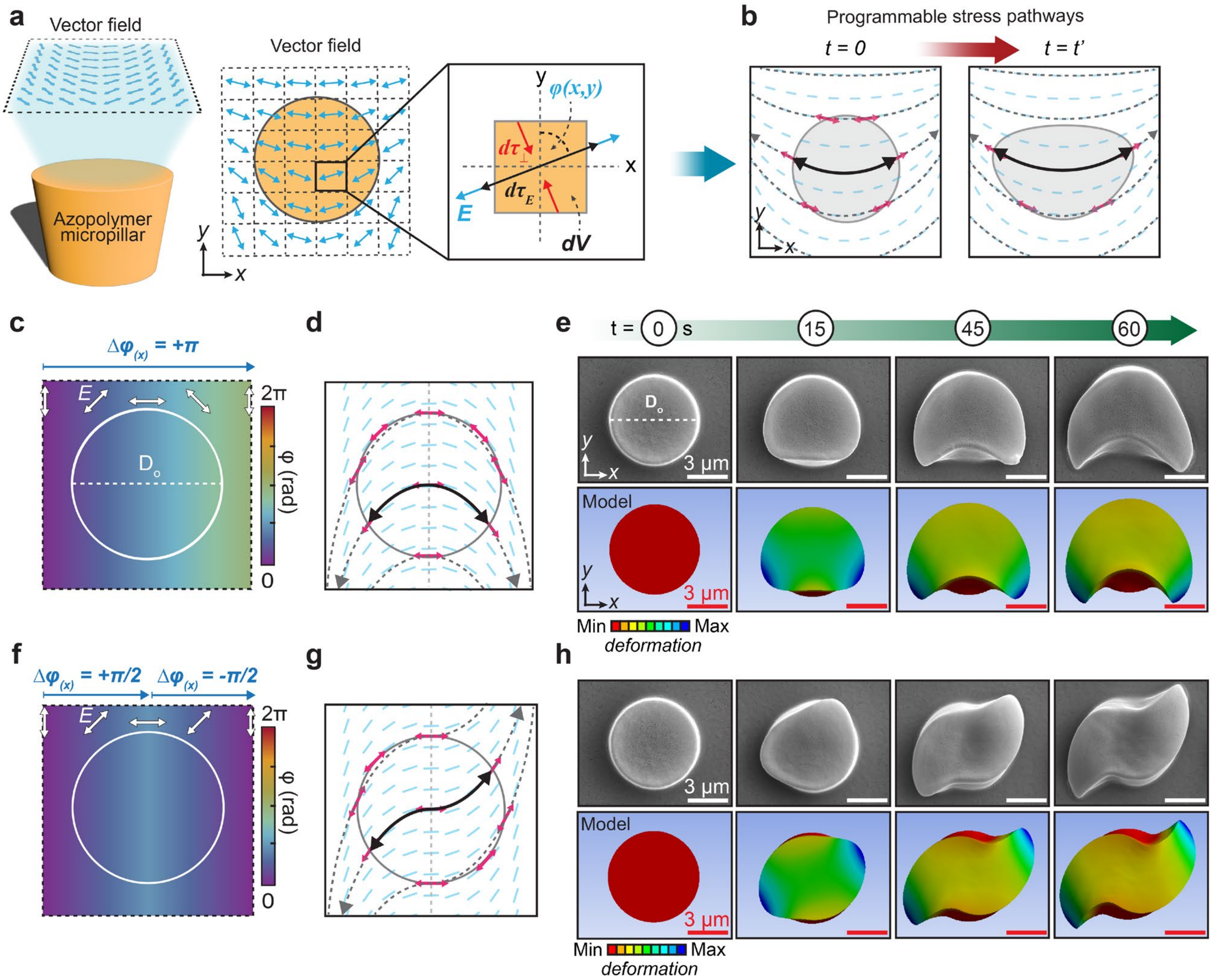

**Fig 4. Vectorial field-guided deformations. a)** Illustration of vector beam illumination on a micropillar and the building mechanism of local stress generated within the micropillar volume. **b)** Programmable stress pathways produced by the vector field in a) at initial and subsequent exposure times. **c)** Polarization map with a $\pi$ rotation of local linear polarization. **d)** Stress pathways scheme emerging from the illumination of the micropillar using the polarization map in c). **e)** SEM (top row) and model (bottom row) images of the morphological evolution of the vectorial field-guided photo-deformation of the micropillar with inverted U-shape. **f)** Polarization map with $+\frac{\pi}{2}$ rotation in the first half of the square map and then with $-\frac{\pi}{2}$. **g)** Stress

pathways scheme emerging from the illumination of the micropillar using the polarization map in f). **h)** SEM (top row) and model (bottom row) images of the morphological evolution of the vectorial field-guided photo-deformation of the micropillar with S-shape. The modeled micropillar is designed to have $D_o$ = 6.7 µm, matching with the experimental micropillar design. The colormap represents vertical displacement: red corresponds to the initial state, while dark blue indicates the maximum downward displacement.

According to this description, the polarization map in Fig. 4c was designed by gradually rotating the polarization azimuth $\varphi$ clockwise from 0 to π along the x-axis of a square segment of $\sim 12 \times 12\ \mu m^2$, while remaining uniform along the y-axis. In the figure, the circular profile of the pristine pillar and its position relative to the designed polarization map is visualized as a white circle of diameter $D_0$. Such a polarization field (see Fig. S11 for the optical characterization) induces an inverted U-shaped stress pathway inside the micropillar volume (Fig. 4d), which consistently guides the deformation of the azopolymer micropillar in the experiment (Fig. 4e, top row). The results of the VPA modeling quantitatively reproduce the observed morphology (Fig. 4e, bottom row) and its temporal dynamics (see also Movie S2). In addition, Fig. 4f shows another square polarization map where the azimuth angle φ is designed to rotate along the x-axis progressively by $\Delta\varphi = +\ \pi/2$ from the left edge to the center and then by $\Delta\varphi = -\pi/2$ from the center to the right edge. This design is apparently similar to the previous case, since it has the same initial and final polarization directions. However, it is characterized by a different relative spatial evolution of the azimuth angle, which generates a chiral stress pathway (Fig. 4g) that leads to the final experimental and modeled S-shaped microstructure shown in Fig. 4h (see also Movie S2).

The striking difference observed in the morphologies of the structures Fig. 4e and Fig. 4h demonstrates the critical role of the induced mechanical stress in the azomaterial photo-reconfiguration process, providing clear evidence of the validity of the VPA model in the general case of structured light irradiation. Not only the local polarization state, but also the spatial evolution dynamics of the polarization ultimately define the actual stress pathway. Structured vectorial light fields then become a powerful tool for guiding the morphological reconfiguration of the microstructures.

Importantly, the polarization maps are not limited to unidirectional uniform linear variations, but can have other spatial dependencies, such as angular evolutions, and may also include discontinuities. Due to the circular profile of our particular pristine micropillars, a relevant example is the polarization maps designed with cylindrical symmetry in a polar coordinate system that uses the micropillar center as the origin of the reference system. In this case, the polarization direction $\varphi$ in the x-y plane depends only on the polar angle $\theta$, while it is uniform along the radial coordinate. Fig. 5a shows a first polarization map designed according to the functional dependence $\varphi(\theta) = \varphi_0 + \theta/2$, where $\varphi_0 = \pi/4$ is the polarization azimuth at $\theta = 0$. This polarization map is expected to generate stress pathways with triaxial symmetry (Fig. S10), whose geometry could also be visually represented in a simplified way by connecting the center of the pillar with the positions at the pillar perimeter where the local polarization is parallel to the local surface normal (the black solid arrows in Fig. 5b).

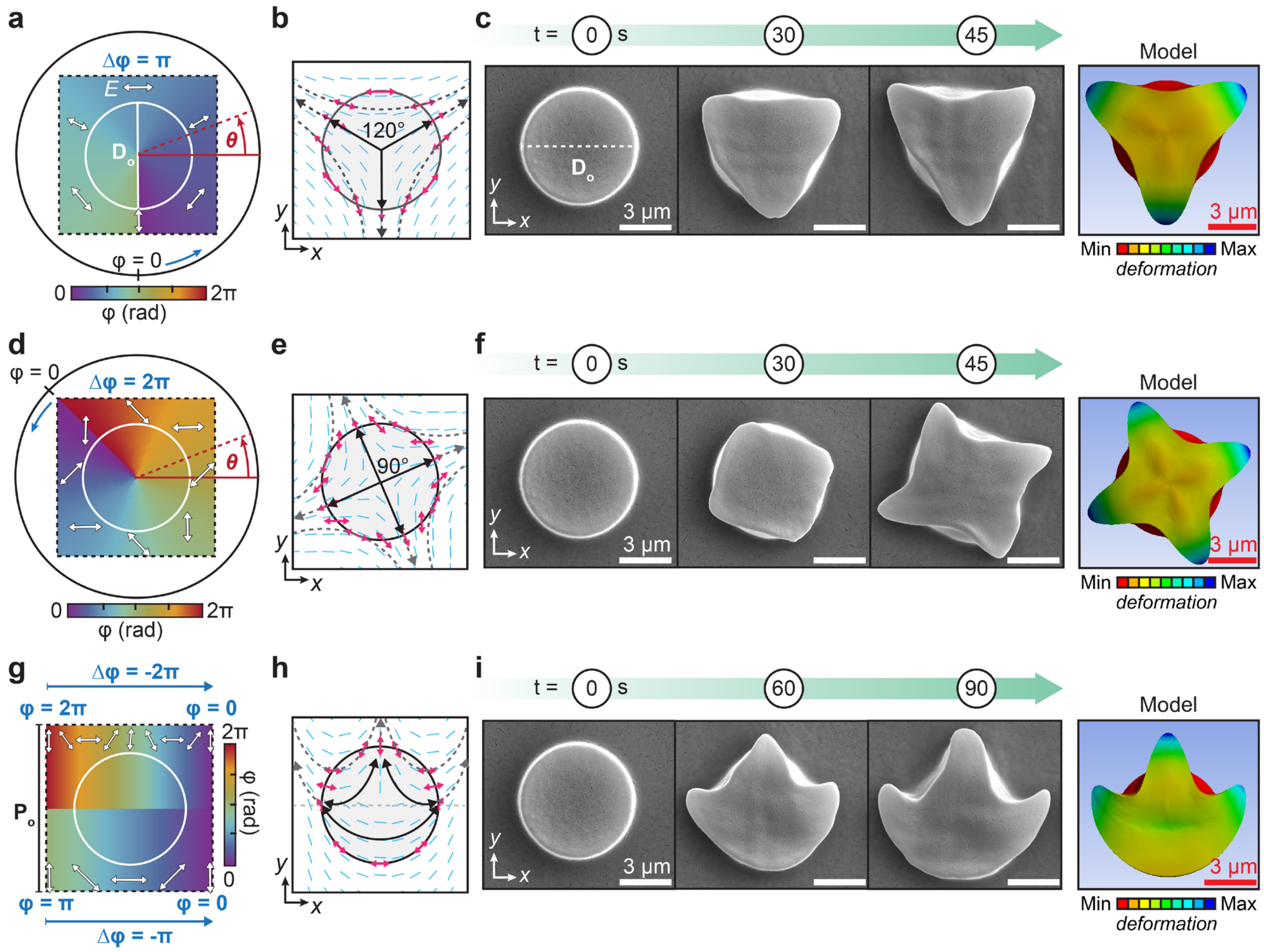

**Fig 5. Design diversity of the vectorial field-guided deformations. a), d)** Polarization maps designed with rotation of azimuth angle following the equation $\varphi(\theta) = \varphi_0 + \theta/2$ , and $\varphi(\theta) = \varphi_0 + \theta$ respectively. $\theta$ is polar coordinate in the x-y plane, with the origin at the center of the micropillar. **b), e)** Stress pathways emerging from a) and d) polarization maps with triaxial and quadriaxial configurations, respectively. **g)** Polarization map with tessellated approach. The top region is designed to undergo $-2\pi$ rotation, the bottom region is designed to undergoes -$\pi$ rotation. **h)** Emerging stress pathways. **c), f), i)** SEM images at different exposure times and model images of the reconfigured micropillars irradiated with the vectorial fields designed in a), d), and g).

Photo-reconfiguration experiments and VPA modeling in Fig. 5c show the expected “tripetal” anisotropic microstructure resulting from irradiation with the designed structured polarization pattern. Notably, such a design is accompanied by a polarization singularity line in the optical field at $\theta = 3\pi/2$ (see also Fig. S11). However, its role in the photo-reconfiguration dynamics was negligible due to the size of the pristine micropillar, which was significantly larger than the region of zero field intensity within the singularity. It is worth mentioning that in ANSYS the polarization rotation is implemented as a local rotation of the element coordinate systems (see Materials and methods) and does not explicitly account for polarization singularities. Accurate treatment of singularities is ensured through the tailored meshing strategy. Then, the agreement of the modeled reconfigured microstructure with the experimental one further confirmed the little influence of the optical singularity in this case. By adjusting the rotation angle of the local linear polarization, the number of stress pathways can be further increased (Fig. 5 d-f). The polarization map shown in Fig. 5d was designed as $\varphi(\theta) = \varphi_0 + \theta$, in which the local linear polarization undergoes a full $2\pi$ rotation along the polar angle $\theta$ and $\varphi_0 = 5\pi/4$. This design creates a quadriaxial system of stress pathways (Fig. 5e), resulting in a photo-reconfigured "quadrupetal" anisotropic microstructure in the model (see also Movie S3), as well as in the experiments (Fig. 5f).

Furthermore, Fig. 5g-i demonstrates the extension of the concept to multiaxial stress-pathways generated by composing multiple polarization maps. The map in Fig. 5g was designed by segmenting the target area into two rectangular tiles with linearly varying polarization patterns. The upper half of the map features a $-2\pi$ rotation of linear polarization from the left to the right edge, while the lower half undergoes a linear $-\pi$ rotation. This polarization field generates three curved stress pathways as shown in Fig. 5h, which deform the microstructure into a peculiar "trident-shaped" anisotropic architecture, correctly described by the VPA model (Fig. 5i and Movie S3).

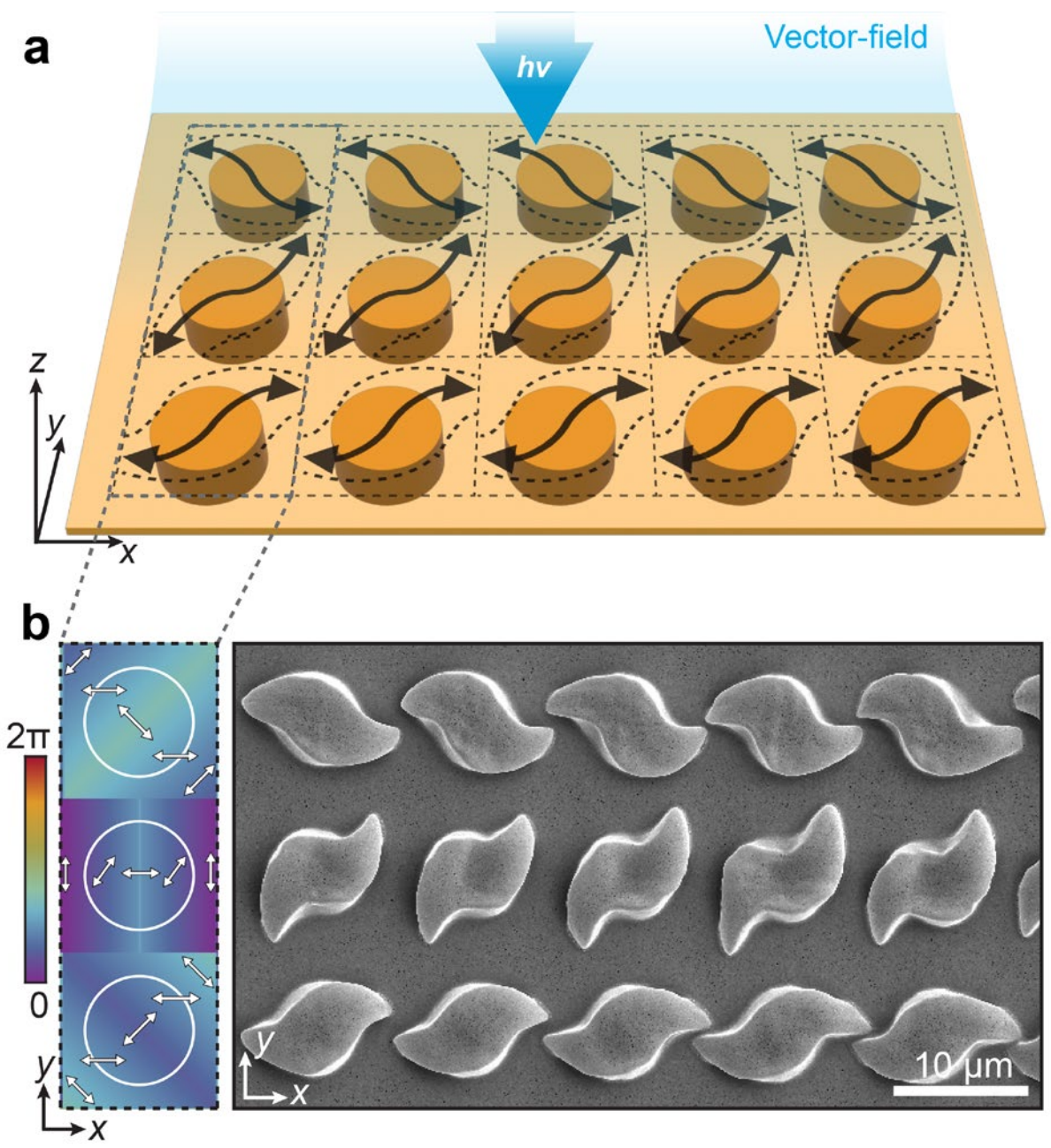

**Fig 6. Design of programmable anisotropic arrays. a)** Illumination scheme of a 5x3 azopolymer micropillar array, featuring spatially selective exposure and locally designed vector fields. **b)** Polarization maps generated starting from the S-shaped architecture and applying three different rotations of the azimuth angle φ (left) to target the five pillars in each row of the array; SEM image of the entire reconfigured 5x3 micropillar array.

As the last step, we demonstrate that the same stress-driven photo-reconfiguration mechanism driving single pillar deformation can be scaled to the entire array fitting into the optical field of view of the projection system (Fig. 6). Given that surface functionality typically arises from the formation of structured arrays rather than from isolated elements, this approach unlocks the potential for creating large-area, multifunctional surface architectures.[47,48]

Fig. 6a schematizes a square-segmented vector field that is aligned with the array of micropillars. Here, the S-shaped geometry shown in Fig. 4f-h was chosen as the base architecture for local pillar reconfiguration in each segment. The segments were then globally rotated by three different angles with respect to the pillar centers, as visible in the polarization maps in Fig. 6b (left column). After exposure, the resulting surface morphologies in the SEM images show the expected array of complex S-shaped microarchitectures with varying global rotations, clearly demonstrating the ability of our vectorial light fields to tailor the shape of surface arrays into desired, on-demand architectures with complex anisotropy.

# Discussion

In this study, we introduce the vectorial field-guided lithography as a new approach for the stress-driven photo-reconfiguration of azopolymer microstructures. By leveraging the core framework of the Viscoplastic PhotoAlignment (VPA) model for the photo-deformation of azopolymers, we demonstrate how structured polarization fields can serve as active and precise tools to control the morphological transformation of azopolymer surfaces. This approach captures the vectorial essence of light and translates it directly into programmable stress pathways that guide the deformation of the microstructures.

Using this approach, we experimentally realize a broad class of complex surface microstructures, having individual or collective anisotropic, bent, and even chiral morphology, through a single-step exposure of a single pre-patterned micropillar array to fully and digitally programmable structured polarization patterns.

Importantly, we validated the VPA model across the full range of structured light conditions, establishing it as a predictive theoretical framework for the light-induced deformation dynamics of azopolymers. This opens, for the first time, inverse design approaches to azopolymer surface morphology, where the structured optical field producing a target morphology can be defined and optimized ahead of the experiment.

Central to our implementation is the digital polarization rotator scheme based on a SLM. While the current setup highlights the power and versatility of the approach, further improvements in spatial resolution and angular polarization control using next-generation modulators (e.g., 10-bit SLMs or high-resolution 4K displays) could extend the method to sub micrometer features. Moreover, the reconfiguration could be scaled to large areas via tiling and motorized positioning, establishing vectorial field-guided lithography as a route to fabricating large-area, multifunctional, surfaces with on-demand anisotropies.

In conclusion, by expanding the role of light from a scalar to a vectorial design parameter, our work introduces a new class of lithography that overcomes intensity-based photopatterning and establishes fully structured light as a tool for optical lithography and surface engineering. The versatility of our approach makes it compatible with any light modulating devices able to address structured light,[49,50] such as SLMs and optical metasurfaces. Notably, when the pre-patterned surface is prepared with higher resolution features (e.g. submicron), even the polarization singularities of the engineered structured field could be included as design parameter in the surface photo-reconfiguration, further extending the set of available optical tools to shape the surface. Potential applications span a wide range of fields, including anisotropic wetting,[30] directional adhesion,[51] surface information encoding,[52] and 3D microstructure-enhanced acoustic streaming.[53]

# Materials and methods

**Azopolymer synthesis and fabrication of the micropillar array**

The azopolymer used in this study was synthesized by radical polymerization of the photosensitive monomer (E)-2-(4-((4-methoxyphenyl) diazenyl)phenoxy)ethyl acrylate, based on the previously reported procedure.[30] Reagents were purchased from Merck and used without further purification. Details of optical absorbance (see also Fig. S1), thermal analysis, and molecular weight distribution have been reported in previous works (Tg = 67 °C, $M_w$=27000 Da).[30,32] The pre-structured film with micropillar array on the glass substrate was fabricated using a soft lithography technique as illustrated in Fig. S2. First, a polydimethylsiloxane (PDMS) soft stamp was fabricated by transferring the surface pattern from a silicone master.[30,38] This process resulted in a PDMS stamp with a negative surface geometry with respect to the target pattern. A 10 wt% azopolymer solution in 1,1,2,2-tetrachloroethane was then deposited onto a standard glass microscope coverslip. The prepared PDMS stamp was then gently placed on the azopolymer solution, allowing the azopolymer solution to infiltrate the negative pattern on the stamp. The sample was then left at room temperature for approximately 5 hours to allow the solvent in the azopolymer solution to completely evaporate. Finally, the PDMS stamp was carefully removed from the azopolymer surface, leaving the pre-structured azopolymer surface on the glass substrate to be used in the experiment. The azopolymer micropillars were designed from different silica masters, with height, $H_o$, diameter, $D_o$, and periodicity, $P_o$, varied according to the experiment being performed.

**Micropillar deformation by homogeneous linearly polarized light**

The experimental study of homogeneous linearly polarized light irradiation was performed using an expanded and collimated 488 nm laser (Coherent OBIS 488 LS) with the intensity at the surface plane set to about 0.16 W/cm$^2$. The pristine micropillars had a measured height of $H_o \approx$ 8.4 μm, diameter of $D_o \approx$ 4.0 μm, and periodicity of $P_o \approx$ 10.0 μm.

**Digital polarization rotator**

The optical setup as shown in Fig. 3b was based on a Coherent Sapphire 488 LP laser that emits a linearly polarized beam at a wavelength of 488 nm. The beam was expanded by two lenses, with focal length $f_1 = -50\ mm$ and $f_2 = 250\ mm$, respectively. This beam expansion increased the beam diameter to uniformly illuminate the entire active area of the Spatial Light Modulator (SLM). In addition, the lenses were arranged to achieve the desired beam waist and divergence for optimal projection. It is important to note that the lens configuration did not form a strictly afocal system (as the distance between lenses is $d_{21} = 380\ mm$) where the image was formed at infinity. Instead, the lenses were positioned to image the light pattern produced by the polarization rotator at a finite distance. This choice minimized the number of optical components in the setup with respect to a conventional $4f$ scheme, contributing to a more compact and efficient system. The polarization rotator unit consisted of a reflective SLM (Meadowlark Optics, Inc., phase-only, 1920 pixels by 1152 pixels), located between crossed quarter-waveplates (QWP I and QWP II). The fast axes of QWP I and II were oriented at +/-45° respectively to input beam polarization direction. This combination comprises a variable phase-retarder, rotating the plane of polarization of a linearly polarized input beam by $d\theta$, where $d\theta$ is set by the programmable optical retardation as determined by the gray value addressed to the SLM. The quantitative characterization of the dependence of the output polarization direction on the input gray value addressed to the SLM was performed by positioning a linear polarizer immediately after QWP II. The results, as shown in Fig. S5, indicate a rotation of 1.4° per grayscale increment, demonstrating that 0 to $2\pi$ radians of rotation can be imparted by each SLM pixel over 0 to 255 grayscale values.

The spatially structured beam was then projected onto the sample plane by a 40× objective with NA = 0.65. The illuminated area corresponded to a demagnified image of the SLM screen with a projected area of approximately 130 μm × 78 μm. The laser intensity used was ~ 7.0 W/cm$^2$. With this demagnification of the SLM screen, the distance between the pixels on the sample plane was approximately 0.07 μm (with a density of $\sim$15 pixels per $\mu m$). This length scale is smaller than the diffraction-limited lateral resolution of the 40×, NA = 0.65 objective at 488 nm, which is ≈0.46 µm. Neighboring SLM pixels therefore oversample the optical point-spread function and allow us to generate a quasi-continuous polarization pattern φ(x, y) that defines the emerging stress pathway.

**Micropillar deformation through the digital polarization rotator**

For collective pillar reshaping (Fig. 3) an array with measured geometrical parameters $H_o \approx 3.0$ µm, $D_o \approx 4.0$ µm, and $P_o \approx 10$ µm was used. The individual micropillar deformation experiment (Fig. 4 - 6) was performed by using an array of azopolymer micropillars with measured geometrical parameters $H_o \approx 3.0$ µm, $D_o \approx 6.7$ µm, and $P_o \approx 12$ µm. To target each micropillar in the array with a $P_o \approx 10$ µm, a box-shaped polarization map with area of 10 × 10 µm$^2$ was used, which corresponds to 148 × 148 pixels in the grayscale image representing the target polarization map. In case of $P_o \approx$ 12 µm, a box-shaped polarization map with area of 12 × 12 µm$^2$ was used corresponding to 178 × 178 pixels in the grayscale polarization design image.

To control the surface plane of the sample in all possible degrees of freedom, a sample holder was placed on an X-Y-Z stage integrated with gimbal mount. The alignment and the light focusing on a specific region of the azo micropillar array was performed by the motorized stages and the gimbals, using a bright field imaging system with a CCD camera as visual feedback.

**Morphological characterization of microstructures**

The morphological characterization of the surface was performed using Scanning Electron Microscopes (Hitachi SU5000, and FEI Nova NanoSEM 450). Prior to image acquisition, the sample underwent a process of metallic coating with a gold layer to increase the conductivity of the surface. The images were obtained using an acceleration voltage of 15 kilovolts and a high vacuum configuration, which is the standard setting for non-conductive samples with a thin conductive layer on the surface in this SEM. The magnifications used for imaging ranged from 1000× to 9000×, depending on the scale of the object of interest.

**Modeling of light-induced stress inside the azopolymer-based microvolume**

Viscoplastic modeling of topographical structures was carried out using the finite element software ANSYS. Specifically, the Perzyna model was employed to describe the relationship between the rate of plastic strain $\dot{\varepsilon}_{pl}$ and the magnitude of light-induced stress $\tau_0$:

$$\dot{\varepsilon}_{pl} = \gamma(\frac{\tau_0}{\tau_{yield}} - 1) \tag{3}$$

where $\tau_{yield}$ represents the yield stress and the parameter $\gamma = \tau_{yield}/(3\eta)$ defines the viscosity $\eta$ of the plastic flow. Under constant light intensity, the ratio $\tau_0/\tau_{yield}$ together with the parameter $\gamma$ determines the timescale of deformation, but not the final morphology. In the present study, $\tau_{yield}$ = 5 MPa was selected based on earlier modeling work[43] and calibrated to reproduce the deformation dynamics of the studied azo-polyacrylate, consistent with expectations that photoinduced processes occur at much lower effective yield stresses than in bulk mechanical tests. The parameter $\gamma$ = 2.5 $*10^{-3}$ $s^{-1}$ cannot be directly measured but was adjusted for the chosen $\tau_{yield}$ to reproduce the experimental timescale of photodeformation. According to Equation (3), the dynamic evolution of photo-deformations is governed by the rate of plastic strain $\dot{\varepsilon}_{pl}$, which determines the speed of the process. This rate increases with the magnitude of the light-induced stress $\tau_0$, which scales with the probability of trans-cis photoisomerization of azobenzene chromophores and thus with the incident light intensity, as predicted by theoretical studies.[15] Consequently, the dynamics can be accelerated by increasing $\tau_0$ and hence the initial tensile component $\tau_E = 2\tau_0/3$. For Fig. 4–5, $\tau_E$ = 25 MPa was chosen to reproduce the experimentally observed deformation profiles under illumination at 7.0 W/cm². For Fig. 2f and S4, where the simulation time was accelerated by a factor of ten, $\tau_E$ = 21.5 MPa was used, corresponding to an intensity of ~0.16 W/cm². The proportionality between $\tau_0$ and the light intensity is expected to hold primarily below ~1 W/cm²; at higher intensities the relation becomes nonlinear, and the model should be regarded as an effective approximation.

The model had a cylindrical shape with different boundary conditions. The bottom surface of the sample was 'glued' to the substrate, restricting both rotational and translational movement. In contrast, the upper and side surfaces were left free to deform. To address the computational challenges while maintaining accuracy, a combination of coarse and fine meshing strategies was employed. Critical regions of the cylindrical geometry, such as the top and bottom surfaces where light-induced deformation occurs, were refined using an element size of 0.25 µm, while less critical areas were discretized with a coarser mesh of 0.5 µm. The finite element mesh was generated using the ANSYS Meshing module, resulting in structured tetrahedral meshes composed of 6862 elements for Fig. 2 and 8733 elements for Figs. 4 and 5, suitable for subsequent finite element analysis.

The light-induced stress tensor was incorporated in ANSYS software using the custom subroutine Userthstrain.[17] The thermal strain was chosen in this subroutine using the Perzyna model (3) such that it produced the plastic strain locally prescribed by the light-induced stress. The principal axes of these two tensors coincide. To implement the polarization rotation, each model element was assigned a local coordinate system with the x-axis aligned with the local linear polarization direction and the z-axis normal to the substrate surface. This procedure directly maps the polarization field $\mathbf{E}(\mathbf{r})$ onto the stress field $\boldsymbol{\tau}(\boldsymbol{r})$.

## Author contributions

Conceptualization: S.L.O., D.McG., M.Sap.

Investigation: IK.J., M.Sal., F.R., N.L., P.P., D.McG., C.P., F.B.

Methodology: IK.J., D.McG., M.Sap., S.L.O.

Software: N.L., P.P.

Resources: S.L.O., F.B., M.Sal., D.McG.

Visualization: IK. J., F.R., M.Sal., N.L.

Supervision: S.L.O.

Writing original draft: IK.J., F.R., M.Sap., S.L.O.

Writing – review & editing: M.Sal., M.Sap., D.McG., S.L.O.

## Data Availability

Source data are publicly available for this paper on Zenodo.org at https://doi.org/10.5281/zenodo.17289199.

## Acknowledgments

S.L.O. and F.R. acknowledge funding from ERC grant (HyperMaSH, 101164874), funded by the European Union. Views and opinions expressed are, however, those of the authors only and do not necessarily reflect those of the European Union or the European Research Council Executive Agency. Neither the European Union nor the granting authority can be held responsible for them. M.Sap. and N.L. acknowledge support by the grant GR 3725/10-1 and P.P. by the project 380321452/GRK2430, both from Deutsche Forschungsgemeinschaft. D.McG. and C.P. acknowledge support from US National Science Foundation (NSF) Division of Electrical, Communications and Cyber Systems, award number 2024118 and 2515653, NSF Division of Materials Research, award number 1919557.

## Conflict of interest

The authors declare no conflicts of interest.

## Supplementary Information

Supplementary information accompanies the manuscript on the Light: Science & Applications website (http://www.nature.com/lsa).

*Supplementary information for*

# Stress-driven photo-reconfiguration of surface microstructures via vectorial field-guided lithography

I Komang Januariyasa,[1] Francesco Reda,[1] Nikolai Liubimtsev,[2] Pawan Patel,[2,3] Cody Pedersen,[4] Fabio Borbone,[5] Marcella Salvatore,[1] Marina Saphiannikova,[2,3,*] David McGee,[4,*] and Stefano Luigi Oscurato[1,*]

[1]Physics Department "E. Pancini", University of Naples Federico II, Complesso Universitario di Monte Sant'Angelo, via Cinthia, 80126, Naples, Italy.

[2]Division Theory of Polymers, Leibniz Institute of Polymer Research Dresden, 01069 Dresden, Germany.

[3]Faculty of Mechanical Science and Engineering, Dresden University of Technology, 01062 Dresden, Germany.

[4] Department of Physics, The College of New Jersey, Ewing, New Jersey 08628, USA.

[5]Department of Chemical Sciences, University of Naples "Federico II", Complesso Universitario di Monte Sant'Angelo, Via Cintia, 80126 Naples, Italy.

[*]Stefano Luigi Oscurato, e-mail: stefanoluigi.oscurato@unina.it

[*]Marina Saphiannikova, e-mail: grenzer@ipfdd.de

[*]David J. McGee, e-mail: mcgeed@tcnj.edu

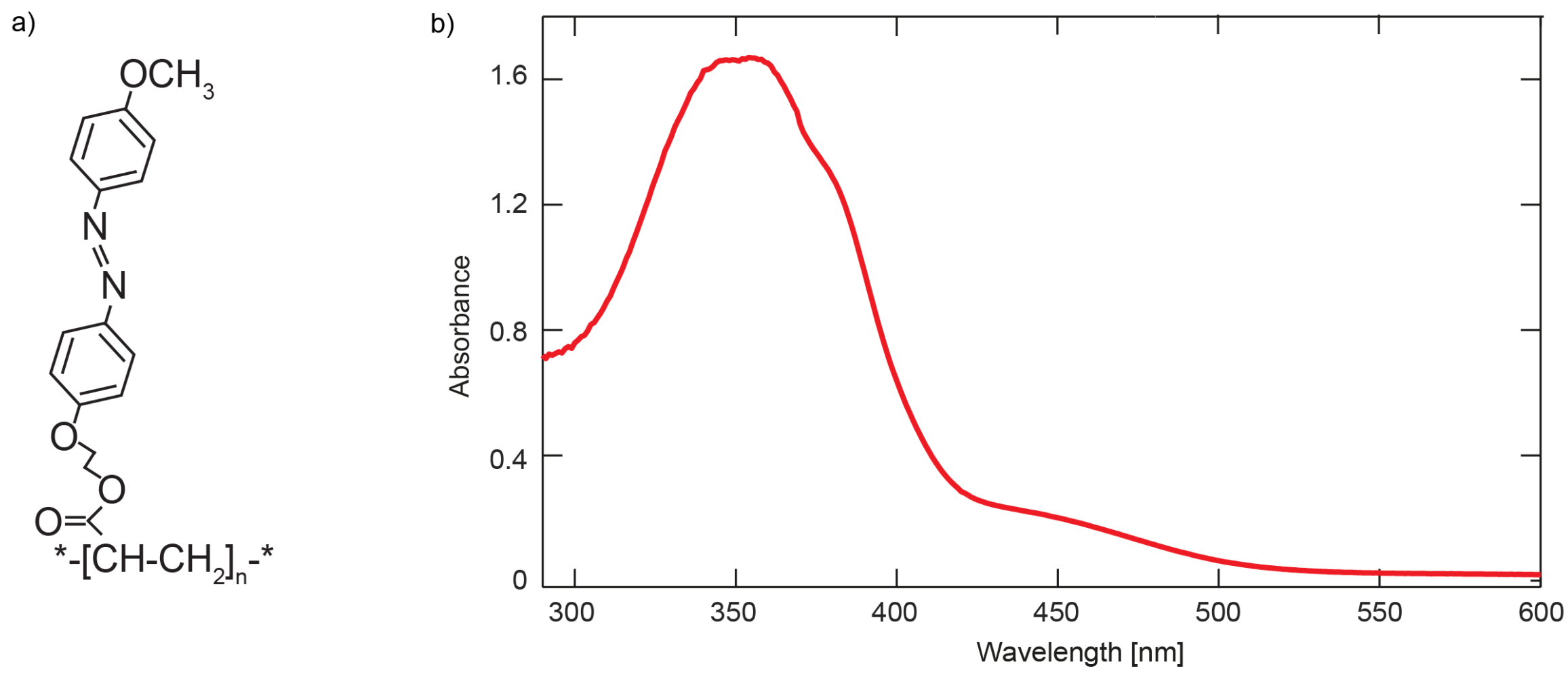

**Figure S1.** a) The chemical structure and b) the UV-Visible absorption spectrum of the azopolymer used in this study, recorded with Jasco V560 spectrophotometer.

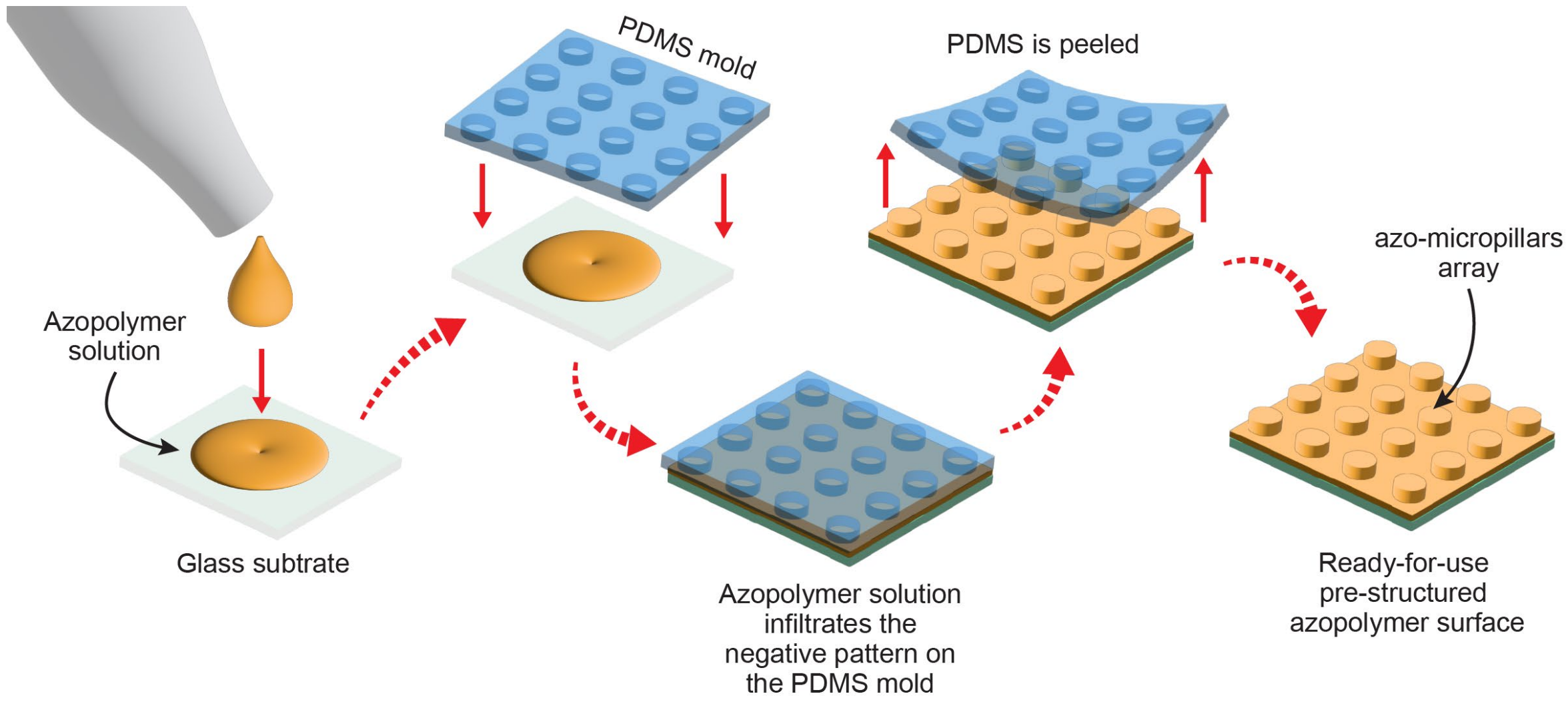

**Figure S2.** The procedure of the soft-lithography used to fabricate azo-micropillars array on the surface. A few drops of 10 wt% azopolymer solution in 1,1,2,2-tetrachloroethane was deposited onto a standard glass microscope coverslip. Then, the prepared PDMS stamp was gently placed onto the azopolymer solution, allowing the azopolymer solution to infiltrate the negative pattern on the stamp. The sample is then left at ambient temperature for approximately 5 hours, during which the solvent in the azopolymer solution evaporates completely. The final step is to carefully peel the PDMS stamp from the azopolymer surface, leaving the pre-structured azopolymer surface on the glass substrate to be used in the experiment. The pre-structured azo-micropillars can be selected to have a height ($H_o$), diameter ($D_o$), and periodicity ($P_o$) which are varied depending on the conducted experiment.

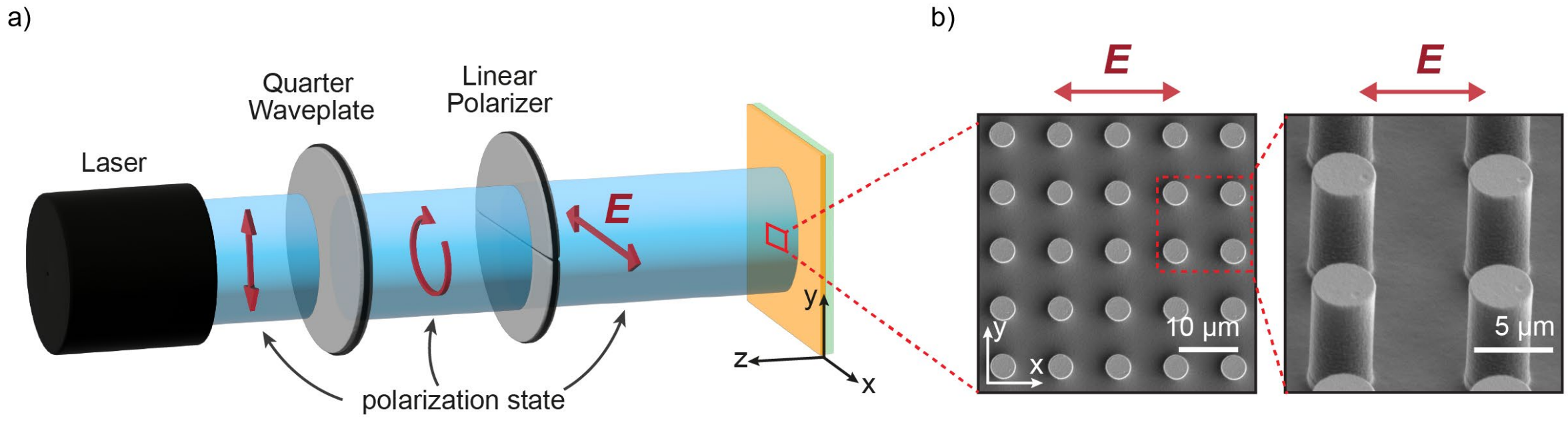

**Figure S3.** The illumination study of homogeneous beam with linear polarization. a) The optical configuration of the illumination exposure. The experimental study of homogeneous linearly polarized light irradiation was performed using a 488 nm laser (Coherent OBIS 488 LS) with the intensity on the surface plane set to about 160 mW cm$^{-2}$. b) The morphology of the pristine state of the azo-micropillar observed from top-down and inclined view. The measured pristine azo-micropillars have a height of $H_o \approx 8.4$ µm, diameter of $D_o \approx 4$ µm, and periodicity of $P_o \approx 10$ µm.

# Supplementary note 1: Viscoplastic PhotoAlignment (VPA) modeling

## 1. Light-induced stress and orientation of backbone segments

The light induced stress $\boldsymbol{\tau}$ in the VPA modeling is defined by the rate of change of the 2nd order orientation tensor $\langle \mathbf{uu} \rangle$ of rigid segments of the polymer backbone:[1]

$$\boldsymbol{\tau} = 3nkT\lambda_{\mathrm{R}} \frac{\partial}{\partial t}\langle \mathbf{uu} \rangle \cdot \tag{S1}$$

Here $\mathbf{u}$ is the unit orientation vector of the segment, $n$ is the number density of segments and $\lambda_{\mathrm{R}}$ is their rotational time in the absence of light. The orientation state of rigid backbone segments is described by the 2nd and 4th order orientation tensors $\langle \mathbf{uu} \rangle$ and $\langle \mathbf{uuuu} \rangle$. The time evolution of 2nd order tensor can be calculated as follows:

$$\frac{\partial}{\partial t}\langle \mathbf{uu} \rangle = -\frac{5V_r}{6\lambda_{\mathrm{R}}}\left\{\hat{\mathbf{E}}\hat{\mathbf{E}} \cdot \langle \mathbf{uu} \rangle + \langle \mathbf{uu} \rangle \cdot \hat{\mathbf{E}}\hat{\mathbf{E}} - 2\hat{\mathbf{E}}\hat{\mathbf{E}} : \langle \mathbf{uuuu} \rangle\right\} - \frac{\langle \mathbf{uu} \rangle}{\lambda_{\mathrm{R}}} + \frac{\boldsymbol{\delta}}{3\lambda_{\mathrm{R}}}, \tag{S2}$$

where $\hat{\mathbf{E}}$ is the unit vector of light polarization and $\boldsymbol{\delta}$ is the unit tensor. The parameter

$$V_r = 2qmV_0 / 5kT \tag{S3}$$

is the reduced strength of orientation potential $U_{\mathrm{eff}}$ acting on the backbone segments irradiated with linearly polarized light:

$$U_{\mathrm{eff}} = qmV_0\left(\hat{\mathbf{E}} \cdot \mathbf{u}\right)^2. \tag{S4}$$

Here $V_0$ is the strength of orientation potential, which effectively describes the gradual orientation of azobenzene chromophores after several photoisomerization cycles. A derivation of effective orientation potential from kinetic equations for the photoisomerization dynamics between trans and cis-isomers of azobenzenes can be found in Ref. 2. Furthermore, it is assumed that each of the backbone segments contains $m$ azobenzenes, whose orientation distribution with respect to the long axes of the segments is described by the shape factor $q$. For side chain azopolymers $q$ is close to -0.5,[3] which means that reorientation of azos perpendicular to the light polarization is transferred to an alignment of the polymer backbones along the same polarization direction. Note that the reduced strength $V_r$ takes negative values because $q < 0$.

## 2. Homogeneous irradiation with linearly polarized light

The side-chain azopolymers align and stretch along the light polarization direction. Let's choose this direction as the axis x of local coordinate system (LCS). Under homogeneous irradiation, i.e. at the constant light intensity, the time evolution of the xx-component of the 2nd order orientation tensor is given by:

$$\frac{\partial}{\partial t}\left\langle u_x^2\right\rangle = \frac{1}{\lambda_{\mathrm{R}}}\left[V_r\left(\left\langle u_x^2\right\rangle - 1\right)\left\langle u_x^2\right\rangle - \left\langle u_x^2\right\rangle + \frac{1}{3}\right] \tag{S5}$$

This expression was derived in Supporting Information to Ref. 4. Due to the axial symmetry with respect to the light polarization, two other diagonal components are equal to each other:

$$\left\langle u_y^2\right\rangle = \left\langle u_z^2\right\rangle = \left(1 - \left\langle u_x^2\right\rangle\right)/2 \tag{S6}$$

and all off-diagonal components are equal to zero. Together with equation (S1), this means that the light induced stress $\boldsymbol{\tau}$ is also diagonal and symmetric around the axis x in LCS:

$$\tau_{yy} = \tau_{zz} = -\tau_{xx}/2\,. \tag{S7}$$

Hence, the tensile component of stress tensor $\tau_E = \tau_{xx}$ and the compressive components $\tau_\perp = -\tau_E/2$ . The time-dependent magnitude of $\tau_E$ is calculated from equation (S1) using equations (S5) and (S6) with the initial value of $\left\langle u_x^2\right\rangle$ = 1/3 which corresponds to isotropic orientation of backbone segments. The rotational time of polymer backbone segments is chosen to be long enough ( $\lambda_{\mathrm{R}} = 1000$ s) to achieve non-vanishing stress values in the VPA modeling of azopolymer micropillars. The reduced potential strength $V_r$ in equation (S5) is prescribed by the magnitude of light induced tensile stress at the beginning of irradiation:[4]

$$\tau_E(0) = -\frac{2}{3} nkTV_r\,. \tag{S8}$$

By implementing the light induced stress tensor into the custom subroutine Userthstrain of the finite element software ANSYS, we are able to reproduce the uniaxial deformation of a homogeneously irradiated azopolymer micropillar. For that we apply the visco-plastic Perzyna model, described by equation (3) in the main text. More details about the implementation of light induced stress can be found in the PhD thesis of B. Yadav.[5]

### 3. Light induced stress at the initial stages of irradiation

The light induced stress in the VPA model is described by rather complicated tensorial equations (S1) and (S2), which contain the dyadic tensor $\hat{\mathbf{E}}\hat{\mathbf{E}}$ and the orientation tensors $\langle \mathbf{uu} \rangle$ and $\langle \mathbf{uuuu} \rangle$. This makes difficult for a non-theorist to comprehend a relation between the light induced stress and the light polarization. Fortunately, it is possible to find an elegant solution to this system of tensor equations, assuming that the initially isotropic state of polymer backbones changes very slowly under light irradiation. Under this condition, the light induced stress remains close to its value at the beginning of irradiation.

First, we immediately see that the last two terms in equation (S2) cancel each other out, since the initially isotropic state is described by $\langle \mathbf{uu} \rangle = \boldsymbol{\delta}/3$. To simplify the expression in parentheses, it is necessary to apply a closure approximation to the 4th order orientation tensor $\langle \mathbf{uuuu} \rangle$. It is known that linear closure provides an exact solution for the random orientation of rods:[6]

$$\langle \mathbf{uuuu} \rangle = -\tfrac{1}{35}\boldsymbol{\Sigma}_4 + \tfrac{1}{7}\langle \mathbf{uu} \rangle \cdot \boldsymbol{\Sigma}_4 + \tfrac{1}{7}\boldsymbol{\Sigma}_4 \cdot \langle \mathbf{uu} \rangle \qquad \text{(S9)}$$

where $\boldsymbol{\Sigma}_4 = \boldsymbol{\delta\delta} + \mathrm{I} + \mathrm{I}^{+}$ represents the sum of $4^{\text{th}}$ order isotropic tensors. Applying the linear closure to the expression in parentheses of equation (S2) results in

$$\hat{\mathbf{E}}\hat{\mathbf{E}} \cdot \langle \mathbf{uu} \rangle + \langle \mathbf{uu} \rangle \cdot \hat{\mathbf{E}}\hat{\mathbf{E}} - 2\hat{\mathbf{E}}\hat{\mathbf{E}} : \langle \mathbf{uuuu} \rangle = \frac{2}{5}\hat{\mathbf{E}}\hat{\mathbf{E}} - \frac{2}{15} Tr(\hat{\mathbf{E}}\hat{\mathbf{E}})\,\boldsymbol{\delta} \qquad \text{(S10)}$$

While the trace of $\hat{\mathbf{E}}\hat{\mathbf{E}}$ is equal to 1 for any polarization state,

$$\frac{\partial}{\partial t}\langle \mathbf{uu} \rangle = -\frac{2}{15\lambda_R}\frac{qmV_0}{kT}\left(\hat{\mathbf{E}}\hat{\mathbf{E}} - \boldsymbol{\delta}/3\right) = -\frac{V_r}{3\lambda_R}\left(\hat{\mathbf{E}}\hat{\mathbf{E}} - \boldsymbol{\delta}/3\right). \qquad \text{(S11)}$$

After substituting the time derivative of $\langle \mathbf{uu} \rangle$ into equation (S1),

we arrive to the equation (1) of the main text:

$$\boldsymbol{\tau} = \tau_0\left(\hat{\mathbf{E}}\hat{\mathbf{E}} - \boldsymbol{\delta}/3\right)$$

where $\tau_0 = -nkTV_r$ is the magnitude of light induced stress at the beginning of irradiation.

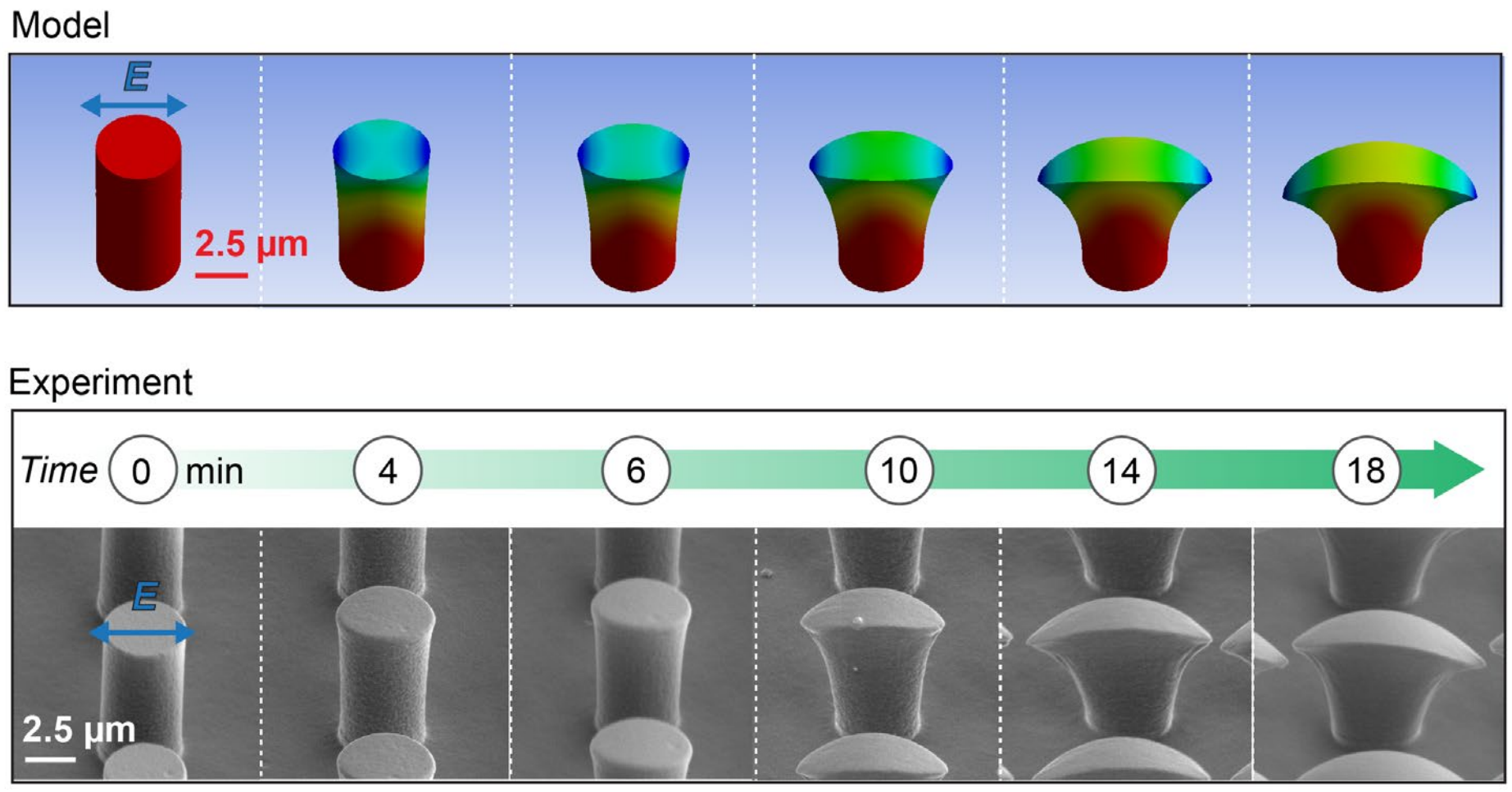

**Figure S4.** The basic study of the azo-micropillar deformation from a homogeneous linearly polarized beam. The morphological evolution of azo-micropillar from the model simulation (top row) and the experiment (bottom row) at different time points (0 – 18 min).

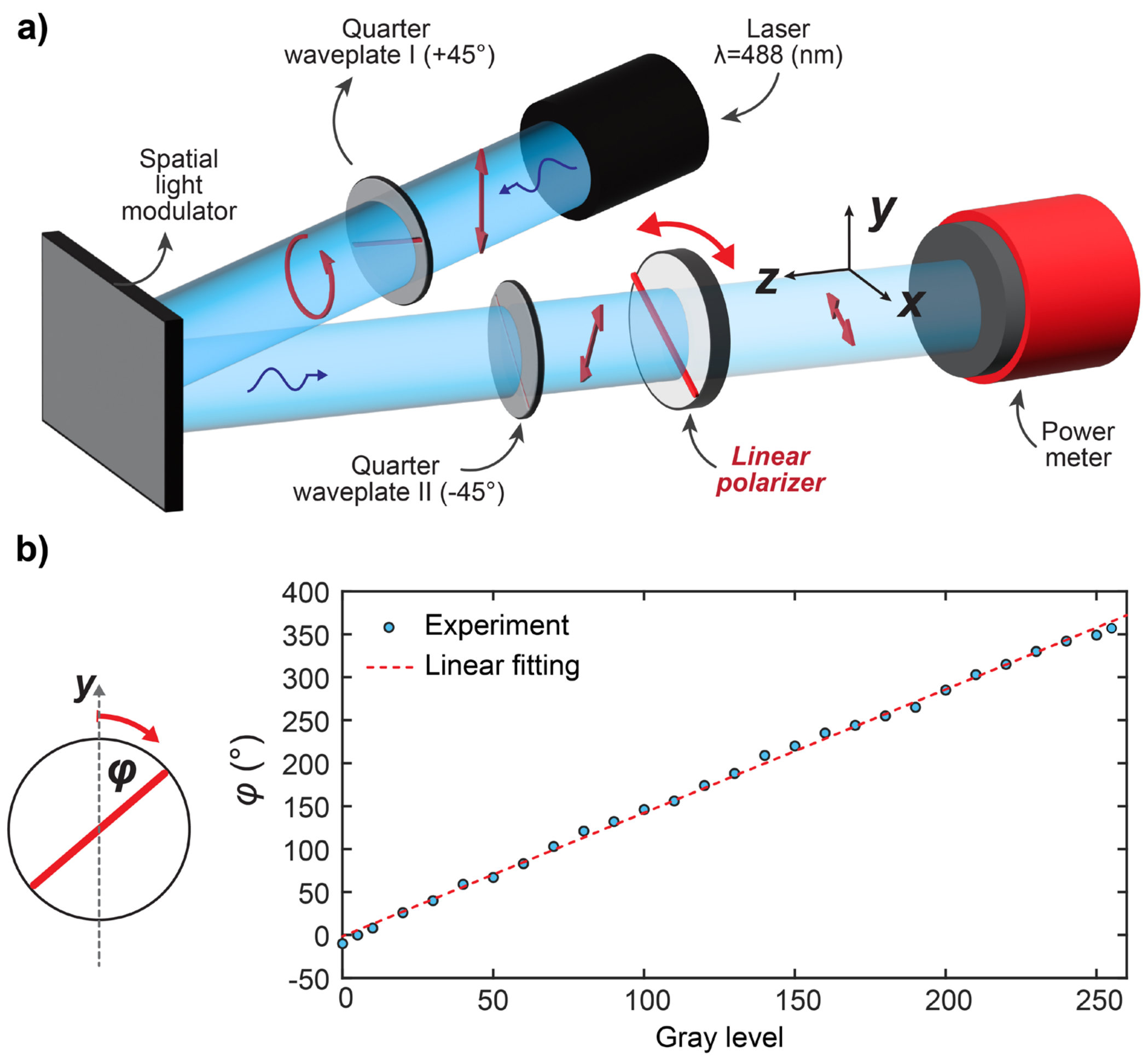

**Figure S5.** a) The optical setup for characterizing the digital polarization rotator. A characterization procedure was conducted to establish a conversion between the grayscale value in the image and the relative rotation of the polarization angle relative to a designated reference axis (the *y*-axis in this study). The procedure of this experiment is to input an image of uniform grayscale level (8-bit, 1920 pixels × 1152 pixels) into the spatial light modulator (SLM), which covered the entire active area, and then to measure the angle of the polarization axis ($\varphi$; relative to the optical *y*-axis) when the power meter measured the maximum power output. This angle represented the major axis of polarization relative to the *y*-axis. The process is repeated for the entire range of grayscale values (i.e., from 0 to 255). b) The curve of conversion grayscale to polarization rotation and its linear fitting.

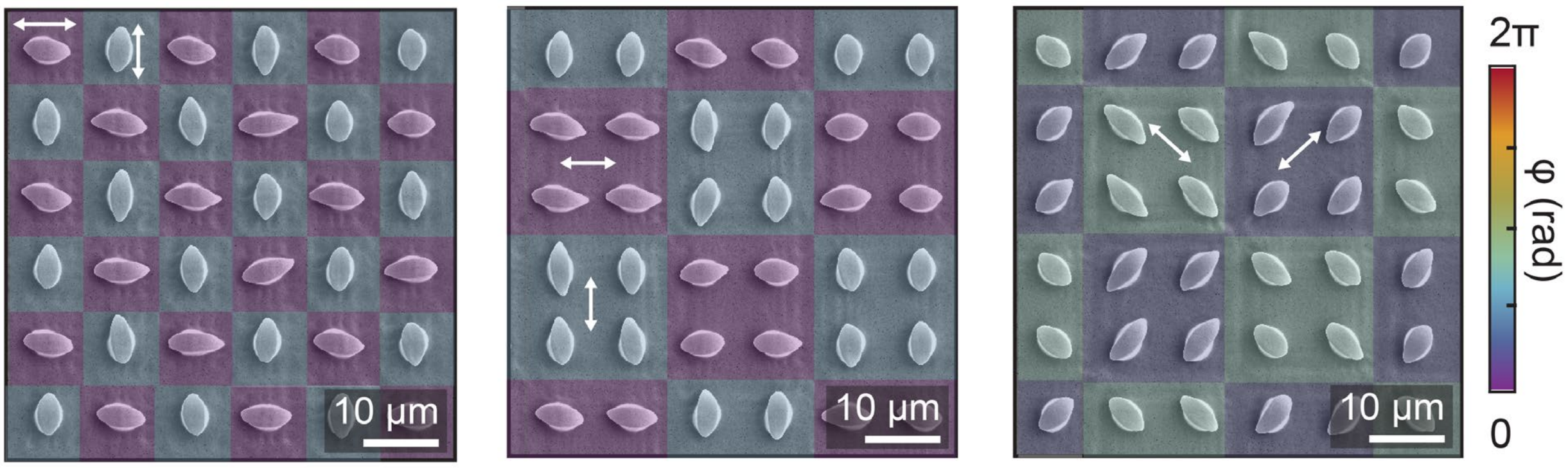

**Figure S6**. Surface reconfigurations with various checkerboard-styled polarization map designs, e.g., vertical-horizontal and diagonal with different numbers of micropillars for each polarization box.

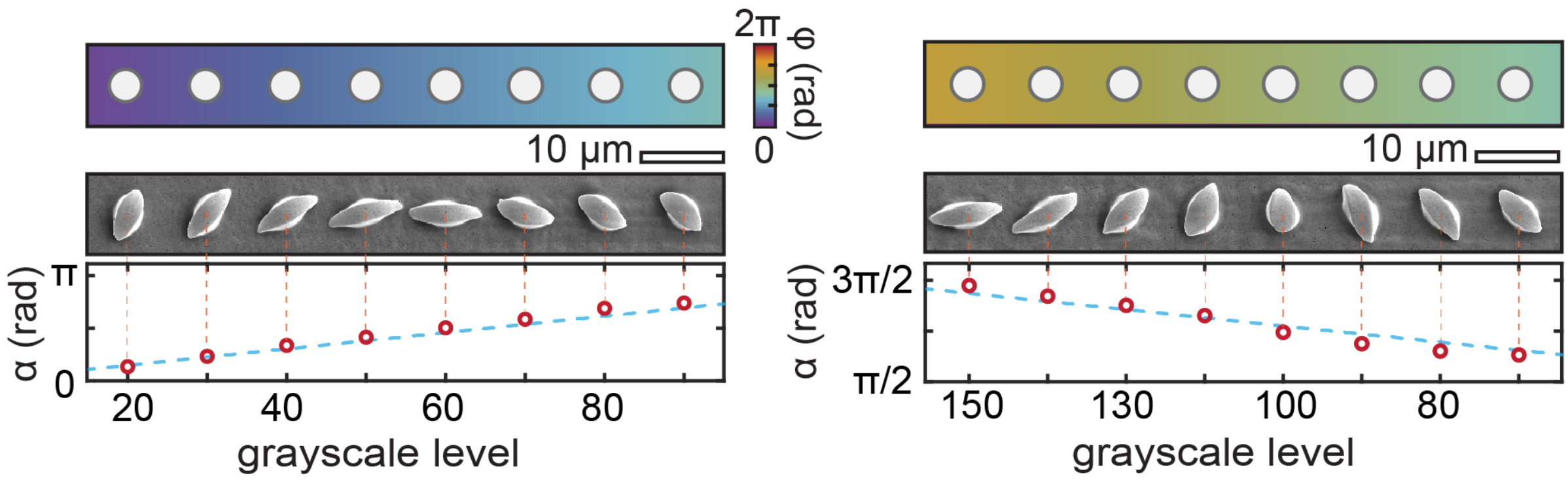

**Figure S7**. Surface reconfigurations with a gradual rotation of polarization azimuth across a row of micropillars. Each reconfiguration was produced with a random grayscale range (e.g., gradually increasing and decreasing trend in left and right data, respectively). α is the angle between the major axis of the micropillar elongation relative to *y*-axis in clockwise rotation. The correlation between the linearly varying polarization pattern and the resulting deformation orientation of the exposed micropillars was studied by measuring the deformation angles α as a function of the linearly increasing gray value assigned to the SLM. The analysis, shown in plot of gray level vs. α, shows that we achieved an excellent agreement with the expected picture of the gradually rotating stress pathways deforming the surface microstructures along the local polarization direction.

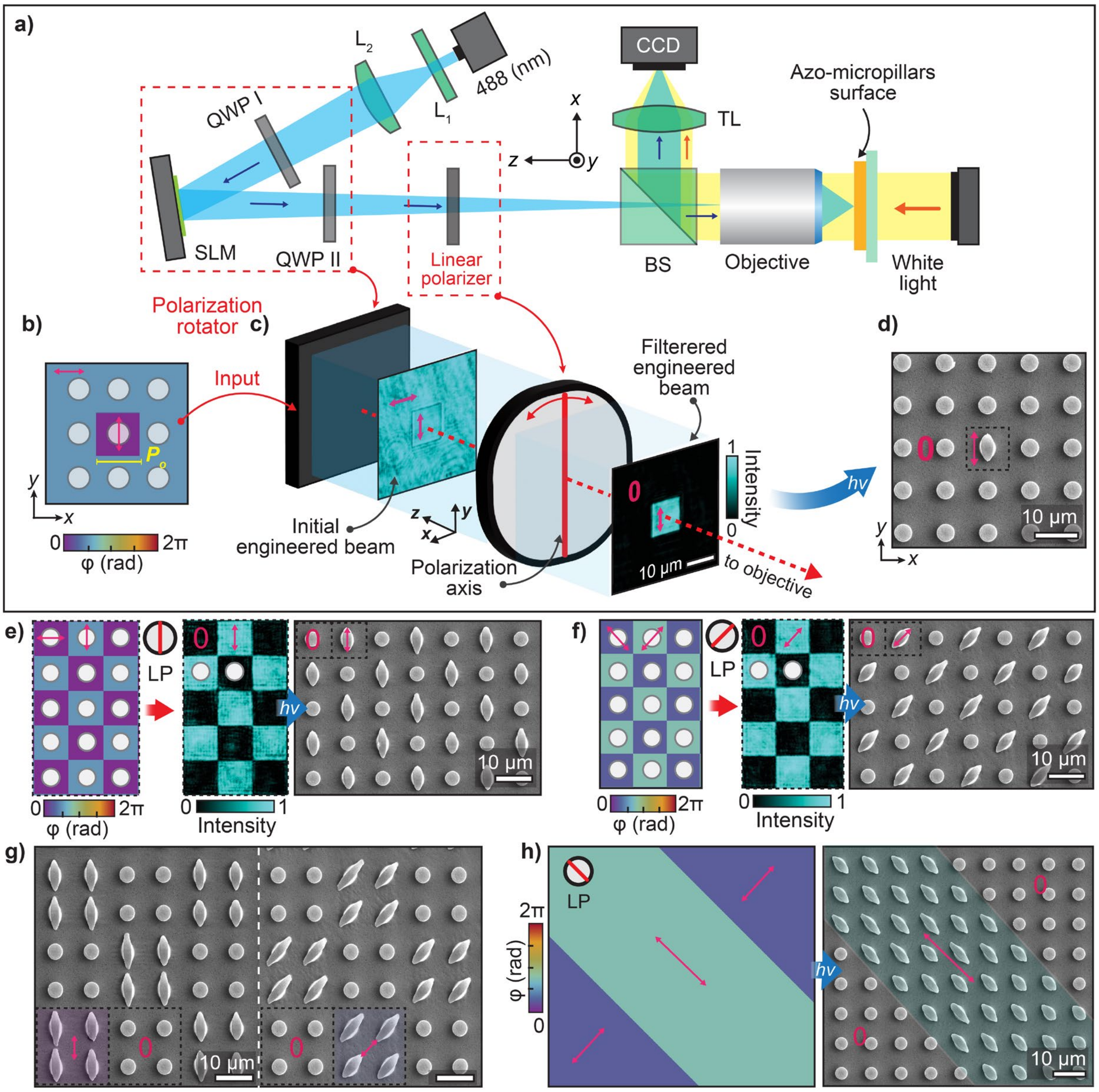

**Figure S8.** The introduction of the "polarization filter" strategy to impart a simple intensity-polarization control in the beam for complex surface configuration. a) The optical configuration of polarization rotator used in this study with an additional linear polarization after the polarization rotator part (i.e., after QWP II). b) The polarization map comprising a rectangular region exhibiting polarization rotation in the vertical direction ($\varphi = 0$), designed to occupy an area encasing a single micropillar ($P_o \times P_o$). The remaining portion of the polarization map is set to possess a polarization rotation in the horizontal direction ($\varphi = 90°$). c) The conceptual workflow of the polarization filter. When the polarization axis of linear polarizer is set to align with the polarization direction of the box region set in (b), it only allows the light of this part to pass and blocks the surrounding beam, producing a box pattern of intensity with vertical polarization direction. This filtered engineered beam provides a spatial selectivity, for example producing d) a deformation only to a single micropillar without affecting the others. The strategy can be expanded to more complex cases, such as a periodic configuration with its own directionality and number of reconfigured pillars (e–g) and an arbitrary pattern (h).

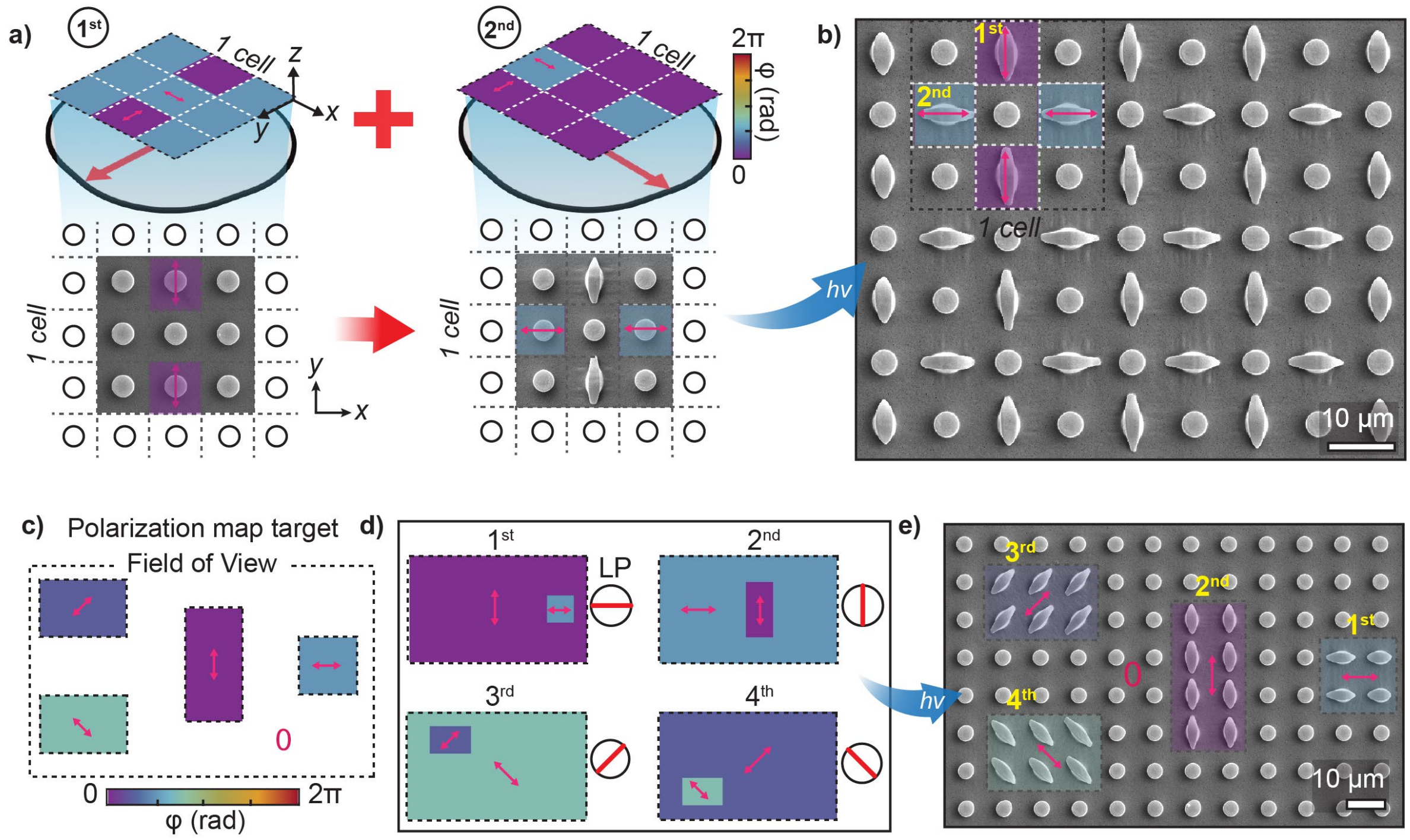

**Figure S9.** The implementation of the "multi-exposure" approach to achieve complex control of spatial selectivity and local directionality in the final surface configuration. a) The conceptual illustration of the 2-exposure approach to produce a periodic quad-petal formation of anisotropic microstructures. The 1st exposure is performed to generate a unidirectional element of the petals (i.e., the petals along the *y*-axis) while leaving an azo micropillar in the center of the petal formation unchanged. The 2nd exposure is then performed to add the other unidirectional element of the petals. b) The final morphology of the quad-petal formation of microstructures array. A demonstration of the multi-exposure approach to produce any configuration of anisotropic surface, such as a) 4-box areas with different sizes and local directionalities. To achieve this pattern, the key is to divide the map target into d) four different exposures. Each exposure is designed to convey one element (i.e., one box) of the target map by designing a polarization map that contains only the intended element of the target map and setting the rest of the field of view to orthogonal polarization. The polarization axis of the linear polarization is then adjusted to filter the engineered beam to have only the intended element of the target polarization. e) The final anisotropic surface configuration that matches the target polarization map.

# Supplementary note 2: Visualization of light-induced stress pathways using ANSYS software

Primary deformation of azopolymer micropillars follows the major principal direction of the maximum tensile component of the photoinduced stress, which is dictated by the local polarization state of the incident light. In Figure S10, the direction of the x-axis in the local coordinate system for each finite element is shown. It is aligned with the major principal axis of the stress tensor, as the incident light field in this case is linearly polarized, producing a consistent spatial distribution of tensile stress pathways.

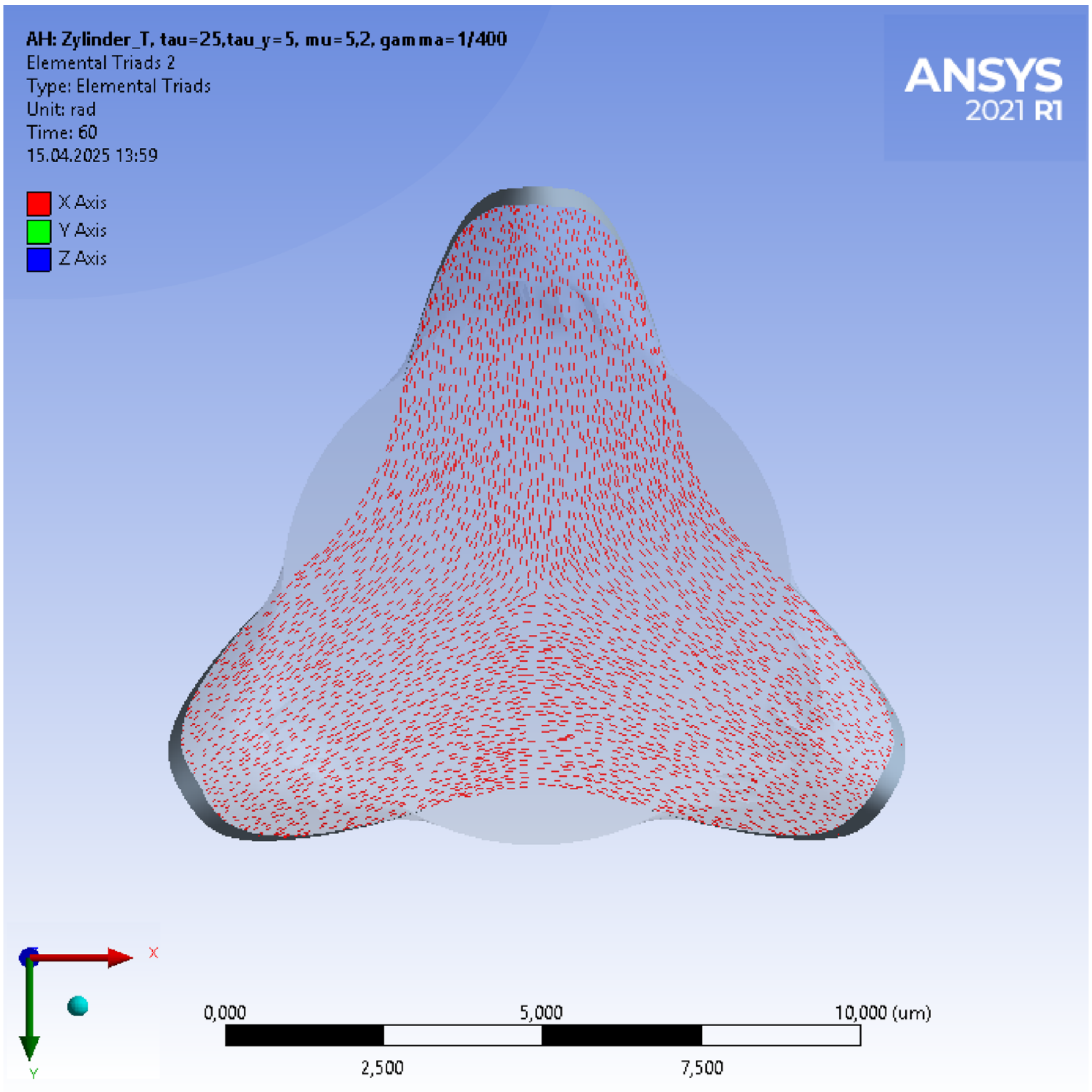

**Figure S10.** Orientation of the x-axis (red arrows) in the local coordinate system for each finite element, aligned with the major principal axis of the stress tensor. The distribution of axes is shown for the triaxial micropillar system.

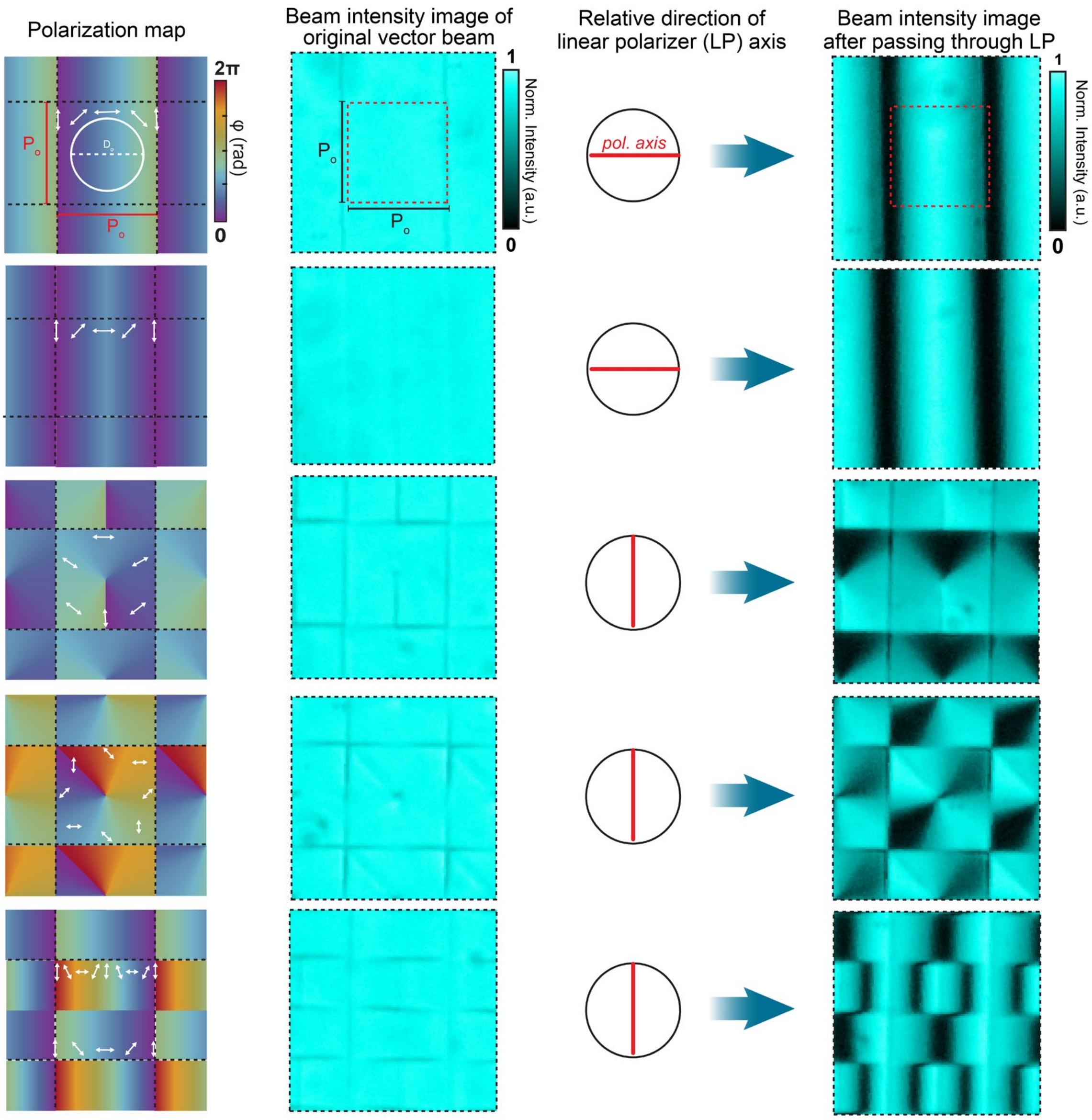

**Figure S11.** The qualitative characterization of the engineered beam with linear polarization filter. In the left column, the white circular line in the polarization map represents the perimeter of a single micropillar with a diameter of $D_o \approx 6.7$ µm. The white arrows provide visual indicators of the local linear polarization, as programmed by the grayscale value addressed to the SLM. To target a single micropillar in a square array, each polarization map is designed to have an area of $P_o \times P_o \approx 12 \times 12$ µm$^2$. The characterization was performed with setup shown in Fig. S9 a. A mirror was placed in the sample plane, and subsequently, the CCD camera was utilized to capture an intensity image that was reflected back from the mirror. The original beam intensity images were taken without a linear polarizer (LP) in front of the CCD camera , as shown in the second column. With an LP as oriented in the third column, the resulting CCD images are shown in the fourth column. In order to improve the clarity of the intensity image description, an LED 477 nm (with full width of half maximum = 29 nm) was used as the light source for this experiment.

**Table S1.** The morphological measurements of different polarization map designs of engineered vector beam.

| **Experiment** | **Initial Diameter $D_o$ (µm)** | **Major axis of elongation $L$ (µm)** | **Minor axis of elongation $l$ (µm)** | **Related figures in main text** |
|---|---|---|---|---|
| Illumination using engineered vector beam with homogeneous azimuth angle of $\varphi = \pi/4$ | $4.0 \pm 0.1$ | $7.5 \pm 0.5$ | $2.5 \pm 0.1$ | Fig. 3e |
| Illumination using engineered vector beam with checkerboard polarization pattern with periodic distribution of $\varphi_1 = \pi/4$ and $\varphi_2 = 3\pi/4$ | $4.0 \pm 0.1$ | $6.3 \pm 0.5$ | $2.7 \pm 0.1$ | Fig. 3f |
| Gradual rotation of local linear polarization along a row of azopolymer micropillars (1st row) | $4.0 \pm 0.1$ | $7.6 \pm 1.1$ | $2.4 \pm 0.2$ | Fig. S7 ($1^{st}$ row) |
| Gradual rotation of local linear polarization along a row of azopolymer micropillars ($2^{nd}$ row) | $4.0 \pm 0.1$ | $7.4 \pm 1.1$ | $2.6 \pm 0.3$ | Fig. S7 ($2^{nd}$ row) |